\documentclass[12pt]{JHEP3}

\usepackage{bbold}
\usepackage{amssymb}
\usepackage{slashed}
\usepackage{graphics}
\usepackage{youngtab}
\usepackage{amsmath}
\usepackage{shadow}
\usepackage{epsfig}
\usepackage{float}

\newcommand{\bm}[1]{\mbox{\boldmath$ #1$}}

\newcommand{\nn}{\nonumber}

\newcommand{\xo}{\stackrel{\circ}{x}\!{}}

\newcommand{\grad}{{\bf grad}}
\newcommand{\g}{{\bf g}}
\renewcommand{\div}{{\bf div}}
\newcommand{\tr}{{\bf tr}}
\newcommand{\N}{{\bf N}}

\newcommand{\eqn}[1]{(\ref{#1})}

\def\bea{\begin{eqnarray}}
\def\eea{\end{eqnarray}}
\def\be{\begin{equation}}
\def\ee{\end{equation}}
\def\ba{\begin{align}}
\def\ea{\end{align}}

\def\sideremark#1{\ifvmode\leavevmode\fi\vadjust{\vbox to0pt{\vss
 \hbox to 0pt{\hskip\hsize\hskip1em
 \vbox{\hsize3cm\tiny\raggedright\pretolerance10000
  \noindent #1\hfill}\hss}\vbox to8pt{\vfil}\vss}}}

\title{Detours and Paths: \\ 
BRST Complexes and Worldline Formalism}

\author{Fiorenzo~Bastianelli$^{\,\clubsuit}$,  Olindo~Corradini$^{\,\clubsuit,\diamondsuit}$  
and Andrew~Waldron$^{\,\spadesuit}$  \\  
$^{\clubsuit}$ Dipartimento  di Fisica, Universit{\`a} di Bologna 
and INFN, Sezione di Bologna\\ $\quad$ via Irnerio 46, I-40126 Bologna, Italy
\\
${}^\diamondsuit$ Laboratoire de Physique Th\'eorique, Universit\'e de
            Paris-Sud XI\\ $\quad$ B\^atiment 210, F-91405 Orsay 
            CEDEX, France\\ 
${}^\spadesuit$ Department of Mathematics, University of California\\ $\quad$ 
            Davis CA 95616, USA\\

\mbox{E-mail: \email{bastianelli,corradini@bo.infn.it},$\ $ \email{wally@math.ucdavis.edu}}}

\abstract{We construct detour complexes from the BRST quantization of worldline diffeomorphism 
invariant systems.  This yields a method to efficiently extract 
physical quantum field theories from particle models with first class constraint algebras.
As an example, we show how to obtain the Maxwell detour complex by gauging 
${\cal N}=2$ supersymmetric quantum mechanics in curved space.
Then we concentrate on first class algebras belonging to a class of recently 
introduced orthosymplectic quantum mechanical models and give generating 
functions for detour complexes describing higher spins of arbitrary symmetry 
types. The first quantized approach  facilitates quantum calculations and we 
employ it  to compute the number of physical degrees of freedom associated 
to the second quantized, field theoretical actions. }


\preprint{LPT Orsay 09-07}

\begin{document}

\section{Introduction}

Unquestionably, gauge theories are a central pillar of modern theoretical physics. 
Although they are usually presented in terms of a local symmetry of a field theoretic  action principle,
it is often useful to describe them in a first quantized language.
The purpose of this Article is to present new tools to analyze gauge field theories  
using a first quantized picture, and to apply them to higher spin theories.
These tools employ an algebraic construction known as the detour complex. The 
basic elements of the detour complex are differential operators 
which form the building blocks of the gauge theory under study.
These differential operators can  be represented as quantum mechanical operators.
Crucially, by forming  first class constraint algebras from them, one may consider {\it worldline} gauge theories.
This procedure gives rise to a particle model which generically is diffeomorphism 
invariant on the worldline, and whose physical spectrum is related to the first quantization of the gauge field
theory. Then, the BRST quantization of the particle naturally provides
a cohomological complex, out of which one builds the detour complex.
Equivalence of the BRST cohomology with the detour cohomology guarantees that the 
correct physical information is properly encoded in the construction.
From the detour complex one identifies an action principle for the gauge invariant 
field equations and may use the associated particle model to extract information about
the quantized version of the theory. 

Our first example---which provides much intuition---is the Maxwell detour.
It describes the abelian gauge theory of differential forms 
and is related to the quantization of the~${\cal N}=2$ spinning particle.
We then consider detour complexes constructed out of the gauging of certain 
orthosymplectic quantum mechanical models~\cite{Hallowell:2007zb}, focusing mostly on symplectic subgroups. 
We use them to analyze the structure of gauge theories for bosonic fields of
higher spins with arbitrary symmetry type~\cite{Labastida:1986gy} (see also~\cite{Labastida:1986ft,Labastida:1987kw,Aulakh:1986cb,DuboisViolette:1999rd,Zinoviev:2002ye};
we refer the reader also to the series of higher spin review Articles~\cite{Bekaert:2005vh}).

Thus, in Section 2 we present the structure and some elements of detour complexes. 
In Section 3 we recall the use of path integrals to treat quantum field theories 
in first quantization.
Section 4 contains the example of the Maxwell detour dealing with the abelian 
gauge symmetries of differential forms. In Section 5 we construct detour 
complexes for higher spin fields of arbitrary symmetry type.
Finally, we present our conclusions in Section 6.

\section{Detour Complexes}

\label{Detours}

As already mentioned, gauge theories are usually presented in terms of a 
local symmetry of an action principle.
Other key ingredients, however, are gauge parameters, gauge fields, field equations
and Bianchi identities. These can all be packaged in a single mathematical
object known as a detour complex. Schematically
\be
\begin{array}{c}
0\longrightarrow 
\scalebox{0.55}{$\left\{\!\begin{array}{c}\mbox{\bf gauge}\\ \mbox{\bf
    parameters}\end{array}\!\right\}$} \stackrel{\bm d} {\longrightarrow}
\scalebox{0.55}{$\left\{\!\begin{array}{c}\mbox{\bf gauge}\\ \mbox{\bf
    fields}\end{array}\!
  \right\}$} 
\rightarrow\cdots
\qquad
\cdots\rightarrow 
\scalebox{.55}{$\left\{\! \begin{array}{c}\mbox{\bf field}\\ \mbox{\bf
      equations}\end{array}\!\right\}$} 
\stackrel{\bm \delta}{\longrightarrow} 
\scalebox{.55}{$\left\{\!\begin{array}{c}\mbox{\bf
    Bianchi}\\ \mbox{\bf identities}\end{array}\!
  \right\}$} 
\longrightarrow0
\\
\ \Big|\hspace{-.8mm}\raisebox{-2.5mm}{\underline{\quad\quad\quad\ \ \ \ \ \raisebox{1mm}{\bm G}\quad\qquad\quad\quad }} \hspace{-1.2mm}{\Big\uparrow}
\end{array}
\ee
where for simplicity in the last two entries we have labeled
the field space with the equations living on that space.
Recall that a sequence of operators is called a complex when consecutive products vanish, so here
\be
{\bm G} {\bm d}= 0 = {\bm \delta \bm G}\, .
\ee
These relations subsume the usual ones of gauge theories. Namely, if~$A$ is a gauge field, then~${\bm G}A=0$ is 
its field equation, while if~${ \alpha}$ is a gauge parameter,~$A\rightarrow A+{\bm d}\alpha$ is the corresponding
gauge transformation. The relation~${\bm  G  \bm d} = 0$ ensures that the field equations are gauge invariant.

We call~${\bm G}$ the detour or long operator since it connects dual complexes. 
Optimally,~${\bm G}$ is self-adjoint, so that one can use it to construct an action 
of the form~$S = \frac12 \int (A, {\bm  G} A)$ where the inner product is the
one following naturally from the underlying first quantized quantum mechanical model. From the action principle 
viewpoint, since an arbitrary variation produces the field equations~${\bm  G} A =0$, specializing to 
a gauge transformation, the operator~$\bm \delta$, dual to~${\bm d}$, must annihilate the equation
of motion~${\bm \delta  \bm G} A =0$. 
This Noether-type identity  generalizes the Bianchi identity for the Einstein tensor 
of general relativity, and is precisely the second relation above.

The simplest example is the Maxwell detour. In that case gauge fields are one-forms, or sections of~$\Lambda^1 M$,
while the gauge parameters are zero-forms~$\Lambda^0 M$. The field equations and Bianchi identities are also 
one and zero-forms, respectively. The differentials are the exterior derivative and codifferential
while  the detour operator is simply their product. In diagrammatic notation
\be
\begin{array}{c}
0\longrightarrow 
\wedge^0M
\stackrel{\bm d} {\longrightarrow}
\wedge^1M
\rightarrow\cdots
\qquad
\cdots\rightarrow 
\wedge^1M
\stackrel{\bm \delta}{\longrightarrow} 
\wedge^0M
\longrightarrow0
\\
\ \Big|\hspace{-.8mm}\raisebox{-2.5mm}{\underline{\quad\quad\quad\ \ \ \ \ \raisebox{1mm}{$\bm \delta \bm d$ }\quad\qquad\quad }} \hspace{-1.2mm}{\Big\uparrow}
\end{array}
\ee
Maxwell's equations are at once recognized as~$\bm \delta \bm d A = 0$, while gauge invariance and the Bianchi identity
follow immediately from nilpotency of~$\bm d$ and~$\bm \delta$. The long operator~$\bm \delta \bm d$ connects the de Rham complex
and its dual. Notice we could also not detour, and continue the de Rham complex, the next entry being
two-forms, or in physics language, curvatures.

There are many other examples of detour complexes, for example: In four dimensions the long operator~$\bm \delta \bm d$ is
conformally invariant. In six dimensions, there exists a higher order, conformally invariant detour
operator~$\bm \delta \Delta  \bm d+\cdots$~\cite{Branson:2003an,Gover}. Another interesting variant is to twist the Maxwell complex by coupling to
the Yang--Mills connection of a vector bundle over~$M$. In this case, one obtains a complex exactly when
the connection obeys the Yang--Mills equations~\cite{Gover:2006ha,Somberg}. In this Article we will concentrate on de Rham detours and 
their generalization to ``symmetric forms''. 
The key idea is to use the relation between first quantized spinning
particles and geometry. We obtain detour complexes by gauging these models and employing BRST
quantization. In particular, the long operator connecting a complex with its dual, corresponds to a
shift in the worldline diffeomorphism ghost number. This representation of the BRST cohomology in terms of
field equations for gauge potentials is achieved by a careful choice of ghost polarizations.
After reviewing the use of path integrals in first quantized approaches, 
we begin by studying generalized Maxwell complexes.

\section{Path Integrals}

Just as for strings, also in particle theory a first quantized approach is often useful. 
While the field theory language is usually appropriate, the
worldline approach can often be applied to more efficiently calculate  quantum corrections,
see \cite{Schubert:2001he} for a review. 
The simplest way to introduce the worldline formalism is perhaps to recall the 
example of a scalar field, whose free propagator and one-loop effective action
can be represented in terms of a quantum mechanical hamiltonian
supplemented by an integration over the Fock-Schwinger proper time.
Much of the heat kernel literature can be classified under this point of view.
An integration over the proper time signals
that one is dealing with the quantum theory of 
a reparametrization invariant particle system;
a relativistic spinless bosonic particle for the scalar field case. 
Similarly a spinning particle with~${\cal N}=1$ supergravity on the 
worldline is related to the quantum theory of a Dirac field \cite{Brink:1976sz}.
More generally~$so(N)$ spinning particles  are related to fields of spin~$N/2$
\cite{Gershun:1979fb}.
Once the connection between reparametrization invariant
particle models and quantum field theories is established, 
it is often advantageous to quantize the mechanical model 
with path integrals, {\it i.e.}, summing over spinning particle worldlines. For example, 
this worldline approach has been used for the~$so(N)$ spinning particle 
in arbitrarily curved spaces with~$N=0,1,2$ to study the effective action
for scalars \cite{Bastianelli:2002fv}, spin 1/2 \cite{Bastianelli:2002qw},
and arbitrary differential forms (including vectors) \cite{Bastianelli:2005vk}
coupled to gravity, respectively. The cases~$N>2$ do not admit 
a coupling to a generic curved space, but in \cite{Bastianelli:2007pv}
the worldline path integral has been considered  in flat space (note that conformally flat backgrounds can  be treated as well~\cite{Bastianelli:2008nm}) where the only information contained in the one-loop effective 
action is the number of circulating physical excitations.

Schematically, to compute the one-loop effective action~$\Gamma$, one 
path integrates over closed worldlines with the topology of the circle~$S^1$. 
Gauge fixing worldline diffeomorphisms produces an integral 
over the proper time~$\beta$  (the circumference of the circle).
In arbitrary dimensions~$D$, the result is
\bea
\Gamma &=& \int_{S^1} {\cal D} X\ e^{-S_{\rm particle}[X]} =
-\frac{1}{2} \int_0^\infty \frac{d \beta}{\beta}\,  
\int \frac{d^D x}{(2 \pi \beta)^{\frac D2}} \, \gamma(x,\beta)
\eea
where the density~$\gamma(x,\beta)$ can typically be computed
in a small~$\beta$ expansion
\be
\gamma(x,\beta)= a_0(x) +  a_1(x) \beta +  a_2(x) \beta^2 + \cdots
\ee
and~$a_n(x)$ are called  heat kernel coefficients.
This expansion applies to massless particles and generically is not convergent  
in the upper~$\beta$ limit, even after renormalization. (Massive particles 
contain an extra factor~$e^{-\frac12 m^2 \beta}$ which improves 
the infrared behavior.)
For a free theory in flat space, the only nonvanishing coefficient is the~$a_0$, 
which is constant and counts the number of physical degrees of freedom 
circulating in the loop. This simplest  of quantum computations is the one
we focus on in this Article.

\section{Maxwell Detour}

\label{Maxwell}

In this Section we derive the Maxwell detour complex described in Section~\ref{Detours}.
Our method relies on BRST quantization and a careful choice of ghost polarizations.
We start by reviewing the underlying supersymmetric quantum mechanical model.

\subsection{${\cal N}=2$ Supersymmetric Quantum Mechanics}

The physical Hilbert space of ~${\cal N}=2$ supersymmetric quantum mechanics  is the space of differential
forms~$\Gamma(\wedge M)$ and, geometrically, its quantized Noether charges are the exterior derivative~$\bm d$, codifferential~$\bm \delta$, form Laplacian~$\Delta_{}$
and the degree operator~${\bf N}$~\cite{Witten:M}. In this Section, we review those results in detail.

The model is described by the action
\be
S=\int dt \Big\{\frac12\, \dot x^\mu g_{\mu\nu}\dot x^\nu 
+ i \bar \psi_\mu \, \frac{\nabla\psi^\mu }{dt}+ 
\frac12\, R_{\mu\nu\rho\sigma}\ \bar\psi^\mu\psi^\nu\bar\psi^\rho\psi^\sigma \Big\}\, ,
\ee
which is invariant under rigid~${\cal N}=2$ supersymmetry ($\varepsilon,\bar\varepsilon)$,
$U(1)$ fermion number symmetry ($\alpha$) and worldline translations ($\xi$)
\bea
\delta x^\mu &=&  i\bar\varepsilon\psi^\mu +i\varepsilon\bar\psi^\mu +\xi\dot x^\mu\, ,
\nn\\[3mm]
{\cal D} \psi^\mu &=& -\varepsilon \dot x^\mu 
 + i\alpha\psi^\mu +\xi\frac{\nabla\psi^\mu}{dt}\, , \qquad
{\cal D} \bar \psi^\mu \ =\  -\bar \varepsilon \dot x^\mu 
 - i\alpha\psi^\mu +\xi\frac{\nabla\bar\psi^\mu}{dt}\, .
\eea
In these formul\ae\ ~${\cal D}$ is the covariant variation:~${\cal D} X^\mu\equiv \delta X^\mu +\Gamma^\mu{}_{\nu\rho}X^\nu\delta x^\rho$ which obviates varying covariantly 
constant quantities. Invariance under supersymmetry follows easily upon noting
the identity
\be
\Big[{\cal D},\frac{\nabla}{dt}\Big] X^\mu = \delta x^\rho \dot x^\sigma R_{\rho\sigma}{}^\mu{}_\nu X^\nu\, ,
\ee
using which leaves only
variations proportional to five fermions that vanish thanks to the second Bianchi identity for the Riemann tensor. 

To quantize the model we work in a first order formulation
\be
\dot x^\mu = \pi^\mu\, ,
\ee
which follows from the action principle
\be
S^{(1)}=\int dt\Big\{p_\mu \dot x^\mu +  i \bar \psi_m \dot \psi^m 
-\frac 12 \pi_\mu g^{\mu\nu} \pi_\nu +
\frac12\, R_{mnrs} \bar\psi^m \psi^n \bar\psi^r \psi^s \Big\}\, . 
\ee
Here we have used the vielbein~$e_\mu{}^m$ to flatten the Lorentz indices on the fermions.
Also the covariant and canonical momenta~$\pi_\mu$ and~$p_\mu$ are related by
\be
\pi_\mu=p_\mu-i \omega_{\mu mn} \bar \psi^m\psi^n\, ,
\ee
where~$\omega_{\mu mn}$ is the spin connection. Since the
symplectic current~$p_\mu dx^\mu +i \bar \psi_m d\psi^m$ is
expressed in Darboux coordinates, we immediately read off the quantum commutation
relations
\be
[p_\mu,x^\nu]=-i\delta^\nu_\mu\, ,\qquad
\{\bar \psi_m,\psi^n\}=\delta^n_m\, .
\ee
Motivated by geometry, we represent this algebra in terms of operators
\be
p_\mu=\frac 1i \frac\partial{\partial x^\mu}\, ,\qquad
\bar \psi_\mu=\frac{\partial}{\partial(dx^\mu)}\, ,\qquad \psi^\mu=dx^\mu\, ,
\ee
acting on wavefunctions
\be
\Psi=\Psi(x,dx)=\sum_{k=0}^D F_{\mu_1\ldots \mu_k}(x)\, dx^{\mu_1}\wedge \ldots \wedge dx^{\mu_k}\, .
\label{wavefunction}
\ee
The variables~$dx^\mu$ are Grassmann (so we will often denote their
wedge products simply by juxtaposition) which means the coefficients
$F_{\mu_1\ldots \mu_k}$ in this expansion are totally antisymmetric,
or in other words differential forms (or sections of~$\wedge^kM$).

The Noether charges corresponding to supersymmetries~$Q,\bar Q$, 
fermion number~$\bm  N$ and worldline translations~$H$ are now operators acting 
on wavefunctions. With a suitable normalization they equal
\bea
Q &=&  i\psi^\mu\pi_\mu\, ,\nn\\[3mm]
\bar Q &=&  i\bar \psi^\mu\pi_\mu\, ,\nn\\[3mm]
\bm  N &=& \psi^\mu \bar \psi_\mu  ,\nn\\[3mm]
H &=& \frac12\pi_m\pi^m-\frac i2 \omega_m{}^{mn}\pi_n-
\frac12\, R_{\mu\nu\rho\sigma}\ \bar\psi^\mu\psi^\nu\bar\psi^\rho\psi^\sigma\, ,
\label{charges}
\eea 
and satisfy a~${\cal N}=2$ superalgebra
\be
\{Q,\bar Q\}=-2 H\, ,\qquad [\bm N,Q]= Q\, ,\qquad [\bm N,\bar Q]= -\bar Q\, 
\ee
with all other (anti)commutators vanishing.
These are quantum results, so their orderings matter and have been carefully arranged
to (i) maintain the classical symmetry algebra and (ii) correspond to
well known geometric operations. (This explains the explicit spin connection appearing
in the Hamiltonian~$H$.) In fact, the charges~$(Q,\bar Q, \bm N,-2H)$ correspond precisely to 
the exterior derivative, codifferential, form degree and form Laplacian acting on differential forms
\be
Q\Psi=\bm d\Psi\, ,\qquad \bar Q\Psi=\bm \delta\Psi\, , \qquad
-2H \Psi = \Delta \Psi\, .
\ee
Indeed the above superalgebra is precisely the usual set of relations for these operators
$$
\bm d\bm \delta+\bm \delta \bm d=\Delta\, , \qquad \bm d\, \bm N=(\bm N-1)\bm d\, ,\qquad \bm \delta \bm N=(\bm N+1)\bm \delta
$$
\be
\bm d\Delta=\Delta \bm d\, ,\qquad \bm \delta \Delta=\Delta \bm \delta\, \qquad \Delta \bm N=\bm N\Delta\, .
\ee
Our next task is to gauge this model and obtain a one-dimensional
supergravity theory whose BRST quantization can then be studied.

\subsection{${\cal N}=2$ Spinning Particle}

In Dirac quantization, gauging the supersymmetry and worldline translation symmetries
of~${\cal N}=2$ supersymmetric quantum mechanics amounts to imposing constraints 
$Q=\bar Q=H=0$.
This is implemented by lapse (alias worldline einbein) and gravitini Lagrange multipliers
$e$,~$\chi$,~$\bar \chi$ in the first order action
\be
S^{(1)}=\int dt \Big\{ p_\mu \dot x^\mu +i \bar\psi_m\dot \psi^m - eH -\bar\chi Q -
\chi \bar Q\Big\}\, ,
\ee
where~$H=\frac12 \pi_\mu\pi^\mu -\frac12\,
 R_{\mu\nu\rho\sigma}\ \bar\psi^\mu\psi^\nu\bar\psi^\rho\psi^\sigma~$,
$Q = i \psi^\mu \pi_\mu$,  and~$\bar Q = i \bar \psi^\mu \pi_\mu$ are now
the classical analogues of~\eqref{charges}. Integrating out the canonical momentum yields
\be
S=\int dt\Big\{
\frac 1{2e} \xo^\mu g_{\mu\nu} \xo^\nu 
+ i \bar \psi_\mu \, \frac{\nabla \psi^\mu}{dt}+ 
\frac e2\, R_{\mu\nu\rho\sigma}\ 
\bar\psi^\mu\psi^\nu\bar\psi^\rho\psi^\sigma 
\Big\}\, ,
\label{action1}
\ee
where the supercovariant tangent vector
\be
\xo^\mu \equiv \dot x^\mu -i\bar \chi \psi^\mu -i\chi \bar \psi^\mu   \, .
\ee
The theory enjoys local supersymmetries
\bea
\delta x^\mu &=& i\bar\varepsilon\psi^\mu +i\varepsilon\bar\psi^\mu  \, ,\qquad
{\cal D} \psi^\mu \ =\ -\frac 1e \xo^\mu \varepsilon \, ,\qquad
{\cal D} \bar \psi^\mu \ = \ -\frac 1e \xo^\mu \bar \varepsilon\, ,\nn\\[2mm]
\delta e &=& 2i\bar\chi\varepsilon  + 2i\chi\bar\varepsilon \, , \qquad\quad
\delta \chi \ = \ \dot\varepsilon \, ,\qquad\quad
\delta \bar \chi \ = \ \dot{\bar \varepsilon}\, .
\label{local-susy}
\eea
Invariance is easily verified by noting that 
${\cal D}\!\xo^\mu = i \bar\varepsilon\frac{\nabla\psi^\mu}{dt}
+i \varepsilon\frac{\nabla\bar\psi^\mu}{dt}
+\frac{\delta e\,  \xo^{\, \mu}}{2e}$.

We must now  BRST quantize the model.
Firstly, notice that its Dirac quantization simply imposes the conditions
\be
\bm d\Psi = \bm \delta \Psi = \Delta \Psi = 0
\ee
on differential forms~$\Psi$. This is the solution to the de Rham cohomology
in terms of divergence-free, harmonic forms.  
These conditions are interpreted as 
the equations of motion for the degrees of freedom propagated by the 
${\cal N}=2$ spinning particle model in the target spacetime. 
They are given in terms of the~$k$-form field strengths~$F_k$ for~$k=0,..,D$ 
of (\ref{wavefunction})
satisfying the Maxwell equations~$\bm dF_k =\bm  \delta F_k=0$. It is well known that the
corresponding number of physical degrees of freedom
is given by~${\rm DoF}= \sum_{k=0}^{D-2} \binom{D-2}{k}= 2^{D-2}$. 
However it is extremely useful to have a gauge theoretic description of these equations, 
for example by introducing auxiliary/gauge fields that can allow for a 
corresponding action principle. 
For this purpose we will employ the more powerful BRST quantization technique.
As we shall see the above equations will correspond to the BRST cohomology at a given 
ghost number.

We proceed in a canonical framework and introduce fermionic worldline diffeomorphism 
ghosts with algebra
\be
\{b,c\}=1\, ,
\ee
along with bosonic superghosts
\be
[p,z^*]=1=[z,p^*]\, .
\ee
Wavefunctions in the BRST Hilbert space now depend also on~$(c,z^*,p^*)$
\be
\Psi=\Psi(x,dx,c,z^*,p^*)=\sum_{s,t=0}^\infty\frac{(z^*)^s(p^*)^t}{s!\ t!}\Big(
\psi_{s,t}+c\, \chi_{s,t}\Big)\, ,\label{BRST wavefunction}
\ee
where both~$\psi_{s,t}$ and~$\chi_{s,t}$ are sections of~$\wedge M$ (ungraded
differential forms). The  choice to represent the BRST Hilbert space in terms
of a Fock space with the above polarization is one of the key points of this Article.
Although, one may {\it suspect} that the choice of polarization does not influence the cohomology of 
the BRST charge~$Q_{\rm BRST}$, as we shall see, it has an extremely important impact on how
that cohomology is {\it represented}. This point was first realized by Siegel, see~\cite{Siegel:1990zf}. In particular, we will find equations of motion expressed in
terms of gauge potentials. These are realized by the the so-called long, or detour operator.
For this it is crucial that we express BRST wavefunctions as an expansion in ghost number that is
unbounded below and above. Only in this way, can we form a detour operator connecting de Rham and dual de Rham
complexes.

On the BRST Hilbert space, it is easy to construct the nilpotent BRST charge, the 
result is
\be
Q_{\rm BRST}=c\Delta+ z^*\bm \delta+ z \bm d -z z^* b\, .
\ee
The first three terms are the ghosts times the constraints while the final term
reflects the first class constraint algebra~$\{\bm d,\bm \delta\}=\Delta$. No further terms
are necessary to ensure
\be
Q_{\rm BRST}^2=0\, ,
\ee
 as this algebra is rank~1.
The other operator we shall need is the ghost number
\be
N_{\rm gh}=cb+z^*p -p^* z\, ,
\ee
which obeys
\be
{}[N_{\rm gh},Q_{\rm BRST}]=Q_{\rm BRST}\, .
\ee
Our task now is to compute the cohomology of~$Q_{\rm BRST}$, namely
\be
Q_{\rm BRST} \Psi = 0 \, ,\qquad \Psi\sim \Psi + Q_{\rm BRST} X\, ,
\ee
which is the topic of the next Section.
 
\subsection{BRST Quantization}

To solve the BRST cohomology,\footnote{We thank Andy Neitzke and Boris
  Pioline for an invaluable collaboration leading to  the results of
  this Section.} we begin by requiring 
that~$\Psi$ is BRST closed. Computing~$Q_{\rm BRST}\Psi$  
we find the following 
conditions on the differential form-valued coefficients
of the BRST wavefunction~\eqref{BRST wavefunction}
\be
\begin{array}{rclcl}
\bm d\psi_{0,t+1}&=&\ \ 0\, ,&\quad&t\geq 0\, ,\\[4mm]
\chi_{s-1,t+1}&=&\bm \delta\psi_{s-1,t}+\frac 1s \bm d\psi_{s,t+1}\,
,&& t\geq0\, ,\quad s\geq1\, ,
\\[4mm]
\Delta\psi_{0,t}&=&\ \bm d\chi_{0,t+1}\, ,&&t\geq 0\, ,
\\[4mm]
\Delta\psi_{s,t}&=&\ s\bm \delta\chi_{s-1,t}+\bm d\chi_{s,t+1}\, ,&&t\geq0\, ,\quad s\geq 1\, .
\end{array}\label{closed}
\ee
The last pair of relations are actually not independent of the first pair save 
for the special case~$t=0$.

We may still shift~$\Psi$ by a BRST exact term~$Q_{\rm BRST} X$, for which we make the 
ansatz
\be
X=\sum_{s,t=0}^\infty\frac{(z^*)^s(p^*)^t}{s!\ t!}\Big(
\alpha_{s,t}+c\, \beta_{s,t}\Big)\, .
\ee
Computing~$Q_{\rm BRST} X$ we find equivalences/gauge invariances
\be
\begin{array}{rclcl}
\psi_{0,t}&\sim&\psi_{0,t}+\bm d\alpha_{0,t+1}\, ,&\quad& t\geq0\, ,\\[3mm]
\psi_{s,t}&\sim&\psi_{s,t}+\bm d\alpha_{s,t+1}+s(\bm \delta\alpha_{s-1,t}-\beta_{s-1,t+1})\, ,&&t\geq 0\, ,\quad s\geq1\, ,\\[5mm]
\chi_{0,t}&\sim&\chi_{0,t}+\Delta\alpha_{0,t}-\bm d\beta_{0,t+1}\, ,&& t\geq0\, ,\\[3mm]
\chi_{s,t}&\sim&\chi_{s,t}+\Delta\alpha_{s,t}-s\bm \delta \beta_{s-1,t}-\bm d\beta_{s,t+1}\, ,&&t\geq0\, ,\quad
s\geq 1\, .
\end{array}\label{equivalence}
\ee
To analyze these equations it helps to invoke the grading by ghost number.
At a given ghost number~$N_{\rm gh}=n$, there are an infinity of form fields
\be
(\psi_{s,s-n},\chi_{s,s-n+1})\, ,
\ee
indexed by~$s$. However, using the closed conditions and exactness
freedom, we can arrange for there to be only a single independent form
field at each ghost number. To see this first examine the second equivalence
relation in~\eqref{equivalence}. By choice of~$\beta_{s-1,t+1}$ we can set 
\be
\psi_{s,t}=0\, ,\qquad t\geq 0\, ,s\geq1 \, .\label{gauge choice}
\ee
Substituting this choice in the closed conditions~\eqref{closed},
we learn that 
\be
\begin{array}{rclcl}
\chi_{s,t}&=&\ \ 0\, ,&& t\geq1\, ,\quad s\geq1\, ,\\[4mm]
\chi_{0,t}&=&\bm \delta \psi_{0,t-1} \, ,&& t\geq1\, ,\\[4mm]
\end{array}\label{closed2}
\ee
so the remaining independent fields are~$\psi_{0,t}$ with~$t\geq 0$ and
$\chi_{s,0}$ with~$s\geq 0$. 
From~\eqref{closed} we see that they obey the closure conditions 
\be
\begin{array}{rclcl}
\bm d\psi_{0,t}\ &=&\ \ 0\, ,&\quad&t\geq 1\, ,\\[4mm]
\bm \delta \bm d\psi_{0,0}&=&\ \ 0 \, ,&&\\[4mm]
\bm \delta \chi_{s,0}\ &=&\ \ 0 \, ,&& s\geq 0\, .\\[4mm]
\end{array}\label{closed3}
\ee
The first and last of these are the closed conditions for the 
de Rham complex and its dual, while the middle relation 
is the detour operator.

Now we study exactness.
Firstly we note that 
\be
\psi_{0,t}\sim\psi_{0,t}+\bm d\alpha_{0,t+1}\, ,\quad t\geq0\, ,
\ee
whose interpretation in terms of de Rham complexes is clear.
Then we observe that maintaining the gauge choice~\eqref{gauge choice}
means that further transformations must obey
\be
\beta_{s,t}=\frac1{s+1} \bm d\alpha_{s+1,t}+\bm \delta \alpha_{s,t-1}\, \quad
t\geq1\, ,\quad s\geq 0\, .\label{compensate}
\ee
Employing this relation,
then from~\eqref{equivalence} and using~$\Delta =\bm d\bm \delta +\bm \delta \bm d$  we have
\be
\chi_{00}\sim  \chi_{00} +\bm  \delta \bm d \alpha_{0,0}+ \bm d(\bm \delta\alpha_{0,0}-\beta_{0,1})
= \chi_{0,0}+\bm \delta\bm  d \alpha_{0,0} + \bm d(\bm d\alpha_{1,1})\, ,
\ee
so that 
\be
\chi_{0,0}\sim \chi_{0,0} + \bm \delta\bm  d\,  \alpha_{0,0}\, .
\ee
Again this matches the detour operator. Finally a similar manipulation for the last equivalence 
in~\eqref{equivalence} for~$t=0$ and~$s\geq 1$ yields
\be
\chi_{s,0}\sim\chi_{s,0}+
\bm \delta(-s\beta_{s-1,0}+\bm d\alpha_{s,0}) - \bm d(\beta_{s,1}-\bm \delta \alpha_{s,0})\, .
\ee
The last term vanishes using~\eqref{compensate}, so calling~$\gamma_s\equiv
-s\beta_{s-1,0}+\bm d\alpha_{s,0}$ ($s\geq1$), we find
\be
\chi_{s,0}\sim\chi_{s,0}+\bm \delta \gamma_s \, ,\qquad s\geq 1\, , 
\ee
which matches perfectly the dual de Rham complex. Therefore we have proven the
equivalence of the BRST cohomology and the Maxwell detour complex
\be
\begin{array}{c}
\cdots
\stackrel{\bm d}{\longrightarrow}
\wedge M
\stackrel{\bm d}{\longrightarrow} 
\wedge M
\stackrel{\bm d} {\longrightarrow}
\wedge M
\qquad
\qquad
\qquad
\wedge M
\stackrel{\bm \delta}{\longrightarrow} 
\wedge M
\stackrel{\bm \delta}{\longrightarrow}
\wedge M
\stackrel{\bm \delta}{\longrightarrow}
\cdots
\\
\ \Big|\hspace{-.8mm}\raisebox{-2.5mm}{\underline{\quad\quad\ \ \ \ \ \raisebox{1mm}{$\bm \delta \bm d$ }\quad\qquad }} \hspace{-1.2mm}{\Big\uparrow}
\end{array}
\ee
The horizontal grading is by ghost number, increasing from left to right. The detour
occurs at ghost number zero, at exactly which point the diffeomorphism ghost number
makes its jump by one unit. In Figure~\ref{snake} we depict the Maxwell detour complex snaking
its way through the BRST Hilbert space.
\FIGURE{\epsfig{file=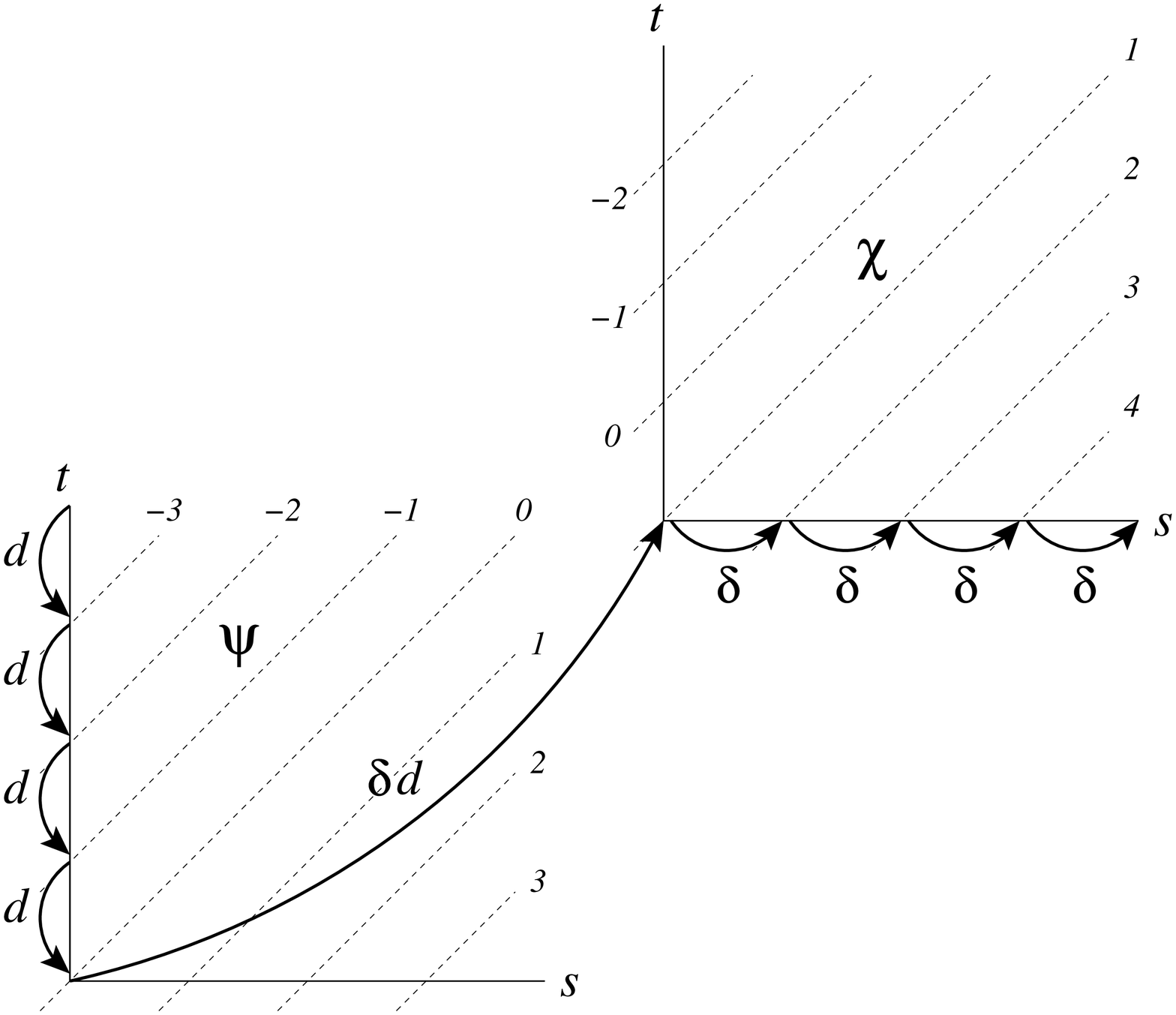,width=9cm}\caption{The Maxwell  complex making its
  way through the BRST Hilbert space. 
Diagonal lines depict form fields of equal ghost number. The first graph
plots fields~$\psi_{s,t}$ and the second~$\chi_{s,t}$.}\label{snake}}

The physical Hilbert space identified by the Dirac quantization procedure
that we summarized earlier on is embedded in the BRST cohomology at 
fixed ghost number ({\it i.e.} zero ghost number for the present case).
As we have seen the same cohomology is reproduced by the detour complex. 
An advantage is that the detour operator which acts at ghost number zero
is formally self adjoint and can be used to construct a field theoretical
gauge invariant action for the degrees of freedom propagated by the 
particle,
that is~$\int A\bm \delta \bm d A \sim \int F^2$ as expected.
The counting of degrees of freedom for this model is well known
from standard work on antisymmetric tensor fields,
and we can reproduce it using the first quantized picture
in a somewhat simpler way.

\subsection{Counting Degrees of Freedom}

To extract quantum  information one can equivalently use either
the first quantized picture of the 
${\cal N}=2$ spinning particle or the second quantized, gauge invariant, field theory action.
The first quantized approach is quite efficient, and we use it  
here to compute the number of physical degrees of freedom. 

We need to evaluate the partition function of the ~${\cal N}=2$ 
spinning particle on the circle to get the one-loop effective action~$\Gamma$. 
With euclidean conventions it reads
\be
\Gamma[g_{\mu\nu}] =
\int_{S^1} \frac{{\cal D} X\, {\cal D} G}{{\rm Vol(Gauge)}} \, e^{-S[X, G;g_{\mu\nu}]}\, ,
\label{1-loop-ea}
\ee
where~$X=(x^\mu,\psi^\mu,\bar \psi^\mu)$ and~$G=(e, \chi, \bar \chi)$ 
indicate the fields that must be integrated over.
$S[X, G;g_{\mu\nu}]$ is the euclidean version of the
action in (\ref{action1}) 
\be
S[X, G;g_{\mu\nu}] =\int_0^1 d\tau\, \Big\{ 
\frac 1{2e} \xo^\mu g_{\mu\nu} \xo^\nu 
+ i \bar \psi_\mu \, \frac{\nabla \psi^\mu}{dt}- 
\frac e2\, R_{\mu\nu\rho\sigma}\ 
\bar\psi^\mu\psi^\nu\bar\psi^\rho\psi^\sigma 
\Big\}\, .
\label{action2}
\ee
The division by the volume of the gauge group implies that we need to fix the 
gauge symmetries. The loop ({\it i.e.} the circle~$S^1$)  is described by taking the 
euclidean time~$\tau \in [0,1]$, imposing periodic boundary conditions 
on the bosons~$(x^\mu,e)$ and antiperiodic boundary conditions on the fermions
$(\psi^\mu,\bar \psi^\mu,\chi,\bar\chi)$.
The gauge symmetries can be used to fix the supergravity multiplet to 
$\hat G=(\beta, 0,0)$, where~$\beta$ is the leftover modulus that must be 
integrated over, {\it i.e.} the proper time.
As the gravitini~$\chi$ and~$\bar\chi$ are antiperiodic, their susy transformations
(\ref{local-susy}) are invertible so that they can be completely gauged away, leaving 
Faddeev-Popov determinants and no moduli. This produces
\bea
\Gamma[g_{\mu\nu}] &=& -\frac{1}{2}
\int_0^\infty \frac{d \beta}{\beta}\,  
\Big ( {\rm {\rm Det}}_{_{A}} \partial_\tau \Big )^{-2} 
\int_{S^1} {\cal D} X\,  e^{-S[X, \hat G;g_{\mu\nu}]}\, ,
\eea
where the proper time measure  takes into account the effect 
of the symmetry generated by the Killing vector on the circle. Note that
the Faddeev-Popov determinants with antiperiodic boundary conditions (denoted by the subscript $A$)
coming from the local supersymmetry do not depend on the target space geometry.
The overall normalization~$(-1/2)$ has been inserted to match with 
the standard result for a single real scalar particle.

We are interested in computing the number of  physical 
degrees of freedom, so we lose no generality\footnote{For partially massless theories~\cite{Deser:2001pe}, more care is needed because
there are various massless limits, but here we are only interested in the strictly massless one.} by taking the flat limit~$g_{\mu\nu} =\delta_{\mu\nu}$,
and evaluate the remaining free path integral over the coordinates~$x^\mu$ 
and their fermionic partner~$\psi^\mu,\bar \psi^\mu$
\bea
\Gamma[\delta_{\mu\nu}] &=& -\frac{1}{2} \int_0^\infty \frac{d \beta}{\beta}\,  
\Big ( {\rm {\rm Det}}_{_{A}} \partial_\tau \Big )^{D-2}
\int \frac{d^D x}{(2 \pi \beta)^{\frac D2}} \ .
\eea
Apart from the standard volume term and the correctly normalized
proper time factors, this result contains the degrees of freedom propagating in the loop,
\be
{\rm DoF} = ( {\rm {\rm Det}}_{_{A}} \partial_\tau  )^{D-2}\, .\ee
The free fermionic determinant is easily computed: 
the antiperiodic boundary conditions produce a trace over the corresponding 
two-dimensional Hilbert space. Thus ${\rm {\rm Det}}_{_{A}} \partial_\tau$  $=$ $2$
and the degrees of freedom~${\rm DoF} = 2^{D-2}$ as expected.
This first quantized picture has been used quite extensively in~\cite{Bastianelli:2005vk} to describe the quantum properties of the 
gauge theory of differential forms coupled to gravity.

\section{Mixed Higher Spin Detour}

In~\cite{Hallowell:2007zb}, supersymmetric quantum mechanical models were constructed with 
$R$-symm\-etries obeying the superalgebra~$osp(Q|2r)$. The ``supercharges'' of these
models transformed under the fundamental representation of~$osp(Q|2r)$,
and therefore generated~$Q$ Grassmann odd supersymmetries and~$2r$ Grassmann
even symmetries. The models were constructed in curved backgrounds.
It was found that the  supercharges commuted with the Hamiltonian only
if the background manifold was a locally symmetric space, at general values of
$(Q,r)$. For low-lying values of~$(Q,r)$ the models coincide
with well known quantum mechanical 
theories. The~$osp(1,0)$ models is the~${\cal N}=1$ supersymmetric quantum mechanics whose single supercharge corresponds
to the Dirac operator and whose Hilbert space describes spinors. The~$osp(2|0)$ theory reproduces the~${\cal N}=2$ supersymmetric quantum mechanics described in Section~\ref{Maxwell}. For both those models the locally symmetric space condition is not necessary.
In this Section we concentrate on the models with~$R$-symmetry~$osp(0|2r)=sp(2r)$. These models are purely bosonic.
Their Hilbert spaces correspond to  multi-symmetric forms and their quantized Noether charges yield symmetrized gradient and divergence type-operators.
We will concentrate on the simplest locally symmetric space--Minkowski space, although many of our computations should generalize easily at least to constant curvature spaces. We begin with the simplest~$sp(2)$ model, which describes totally symmetric tensors or ``symmetric forms''.

\subsection{Symmetric Forms and~$sp(2)$ Quantum Mechanics}

Symmetric forms share many similarities with their totally anti-symmetric counter\-parts--differential forms.
They are expressed in terms of totally symmetric tensors and {\it commuting} coordinate differentials so that 
a symmetric rank~$s$ tensor~$\varphi_{(\mu_1\ldots \mu_s)}$ becomes
\be
\Phi = \varphi_{\mu_1\ldots \mu_s} dx^{\mu_1}\cdots dx^{\mu_s}\, .
\ee 
There is an algebra of operations---gradient, divergence, metric, trace and the wave operator first introduced by Lichnerowicz~\cite{Lichnerowicz:1964zz} and
systemized in~\cite{Damour:1987vm,Hallowell:2005np,Hallowell:2007zb} (see~\cite{Labastida:1987kw,Vasiliev:1988xc,Duval,Duval1} for other studies) ---
which greatly facilitates computations when the rank~$s$ is large or even arbitrary. In particular, it is important to note
that in  this algebra (just as for differential forms)  it is no longer forbidden to add tensors of different ranks.

In flat space,  there are six distinguished operators mapping symmetric tensors to symmetric tensors:
\begin{enumerate}
\item[${\bf N}$\;] --Counts  the number of indices \be {\bf N}\ \Phi = s \Phi \, .\ee
\item[${\bf tr}$\ ] --Traces over  a pair of indices \be {\bf tr}\  \Phi = s(s-1)\varphi^\rho{}_{\rho\mu_3\ldots \mu_s} dx^{\mu_3}\cdots dx^{\mu_s}\, .\ee
\item[${\bf g}$\;] --Adds a pair of indices using the metric
\be {\bf g}\  \Phi = g_{\mu_1\mu_2}\varphi_{\mu_3\ldots \mu_{s+2}} dx^{\mu_1}\cdots dx^{\mu_{s+2}}\, .\ee
\item[${\bf div}$\ ] --The symmetrized divergence 
\be
{\bf div} \ \Phi = s\nabla^\rho \varphi_{\rho\mu_2\ldots \mu_s} dx^{\mu_2}\cdots dx^{\mu_s}\, .
\ee
\item[${\bf grad}$] --The symmetrized gradient
\be {\bf grad} \ \Phi = \nabla_{\mu_1}\varphi_{\mu_2\ldots \mu_{s+1}} dx^{\mu_1}\cdots dx^{\mu_{s+1}}\, .\ee
\item[$\Delta$] --The Bochner Laplacian 
\be
\Delta=
\nabla^{\mu} \nabla_{\mu}\, .\qquad\qquad
\ee
\end{enumerate}
The calculational advantage of these operators is the algebra they obey
\bea
[{\bf N},{\bf tr}]=-2{\bf tr}\, ,\quad
[{\bf N},{\bf div}]=-{\bf div}\, ,\quad
[{\bf N},{\bf grad}]={\bf grad}\, ,\quad
[{\bf N},{\bf g}]=2{\bf g}\, ,\nn
\eea
\bea
[{\bf tr},{\bf grad}]=2{\bf div}\, ,\quad
[{\bf tr},{\bf g}]=4{\bf N}+2D\, ,\quad [{\bf div,g}]=2{\bf grad}\, ,\nn\eea
\bea
[{\bf div,grad}]=\Delta\, .\label{gold}
 \eea
All other commutators vanish. 

Symmetric forms may be interpreted as the Hilbert space of a quantum mechanical model
whose quantum Noether charges are given by the operators above~\cite{Hallowell:2007zb}.
For flat backgrounds this theory is described by the simple action
\be
S=\int dt\left\{\frac12 \dot x^{\mu}\dot x_{\mu}+i z^{*}_{\mu}\dot z^{\mu}\right\}\, .\label{qm}
\ee 
Here the complex variables~$(z^{\mu},z_{\mu}^{*})$ are viewed as oscillator degrees of freedom
describing the index-structure of wave functions so that upon quantization
\be
z^{*\mu}\mapsto dx^{\mu}\, ,\qquad z_{\mu}\mapsto \frac{\partial}{\partial(dx^{\mu})}\, .
\ee
The model's symmetries, Noether charges and their relation to the geometric operators given above is
described in detail in~\cite{Hallowell:2007zb}. In particular, the~$sp(2)$ ``$R$-symmetry''
is generated by the triple~$({\bf tr}, {\bf N}+\frac D2, {\bf g})$. The pair of operators~$\{{\bf div, grad}\}$ transform
as a doublet under this~$sp(2)$ and their commutator produces  the Hamiltonian which corresponds to the Laplacian~$\Delta$.
In this sense,~$\{{\bf div,  grad}\}$ could be viewed as pair of ``bosonic supercharges''.

Totally symmetric tensor higher spin theories can be formulated in terms of the algebra~\eqn{gold}, so this relationship
between those operators and the quantum mechanical model~\eqn{qm} yields a first-quantized worldline approach to these
models. From that viewpoint we need to construct a spinning particle model by gauging an appropriate set of symmetries.
The BRST cohomology of that one dimensional gauge theory then yields the physical spectrum of a higher spin field theory.
We construct this spinning particle model in Section~\ref{sp2rspin}. When the symmetries that we choose to gauge 
form a Lie algebra, the BRST problem becomes equivalent to one in Lie algebra cohomology. This is the topic of our next Section.

\subsection{BRST Quantization and Lie Algebra Cohomology}

In this Section we formulate the theory of massless, totally  
symmetric higher spins~\cite{Fronsdal:1978rb,Curtright:1979uz} as the Lie algebra cohomology of a very simple algebra~$\mathfrak g$:
\be
{\mathfrak g}_{A}=\Big\{{\bf tr}\, , \;
{\bf div}\, , \;
{\bf grad}\, , \;
{\Delta}\Big\}\, ,\label{that one}
\ee
acting on the vector space~$V$ of symmetric forms. The BRST quantization for this algebra was first studied in~\cite{Beng}
(see also~\cite{Tsu,BARN}).

Let us very briefly review the relationship between BRST quantization and Lie algebra cohomology.
In BRST quantization the BRST Hilbert space is expressed as wavefunctions expanded in anticommuting ghosts
\be
\Psi_{\rm BRST} = \sum_{k=0}^{\rm dim\mathfrak g} c^{A_{1}}\cdots c^{A_{k}} \Psi_{A_{1}\ldots A_{k}}\, ,
\ee
while in Lie algebra cohomology the~$V$-valued wavefunctions~$ \Psi_{A_{1}\ldots A_{k}}$ are viewed as 
multilinear maps~${\mathfrak g}^{\wedge k}\rightarrow V$ and form the cochains of a complex. 
The cochain degree~$k$ is BRST ghost number.
The Chevalley--Eilenberg
differential~$\delta$~\cite{Chevrolet} 
\be
\Psi_{A_{1}\ldots A_{k}} \stackrel\delta\rightarrow {\mathfrak
  g}_{[A_{1}} \Psi_{A_{2}\ldots A_{k+1}]}- \frac k2 f^{B}_{[A_{1}A_{2}}\Psi_{|B|A_{3}\ldots A_{k+1}]}\, ,
\ee
can be compactly expressed in terms of the BRST charge
\be
Q_{\rm BRST}= c^{A}{\mathfrak g}_{A} - \frac12 f^{C}_{AB}c^{A}c^{B}\frac{\partial}{\partial c^{C}}\, ,
\ee
acting at ghost number~$k$. The cohomology of~$Q_{\rm BRST}$ at ghost number~$k$ equals the Lie algebra cohomology~$H^{k}({\mathfrak g},V)$.

Returning to our specific Lie Algebra~\eqn{that one}, we now relate its Lie algebra cohomology at degree one to the massless higher spin theory.
At degree one, our problem is a very simple one: We first  introduce a wavefunction for every generator
\be
\Psi_{A}=\Big\{\Psi_{\bf tr},\Psi_{\bf div},\Psi_{\bf grad},\Psi_{\rm \Delta}\Big\}\, .
\ee
The closure condition~$\Psi_{A}\in {\rm ker} \delta$ yields a set
of~$\binom{{\rm dim \mathfrak g}}{2}$ differential equations   
following directly from the commutation relations~\eqn{gold}
\bea
{\bf tr}\ \Psi_{\bf div}-{\bf div}\ \Psi_{\bf tr}\; &=&\;\;0\, ,\nn\\[2mm]
{\bf tr}\ \Psi_{\bf grad}-{\bf grad} \ \Psi_{\bf tr}&=&2\, \Psi_{\bf div}\, ,\nn\\[2mm]
{\bf tr}\ \Psi_{ \Delta}-{ \Delta}\  \Psi_{\bf tr}\; \; &=&\;\;0\, ,\nn\\[2mm]
{\bf div}\ \Psi_{\bf grad}-{\bf grad}\ \Psi_{\bf div}&=&\;\Psi_{\Delta}\, ,\nn\\[2mm]
{\bf div}\ \Psi_{\Delta}-{\Delta}\ \Psi_{\bf div}\; &=&\;\;0\, ,\nn\\[2mm]
{\bf grad} \ \Psi_{\Delta}-{\Delta} \ \Psi_{\bf grad}&=&\;\;0\, .\label{pde}
\eea
Exactness,~$\Psi_{A}\sim \Psi_{A} + X_{A}$ with~$X_{A}\in {\rm im}\delta$, yields the gauge invariances
of this set of equations
\bea
\delta \Psi_{\bf tr}&=&{\bf tr}\ \xi\, ,\nn\\[2mm]
\delta \Psi_{\bf div}&=&{\bf div}\ \xi\, ,\nn\\[2mm]
\delta \Psi_{\bf grad}&=&{\bf grad}\ \xi\, ,\nn\\[2mm]
\delta \Psi_{\Delta}&=&{\Delta}\ \xi\, .
\eea
The fields $(\Psi_\Delta,\Psi_{\bf grad},\Psi_{\bf div})$ correspond to the BRST triplet structure discussed in~\cite{Tsu}
(denoted $(C,\varphi,D)$) while $\Psi_{\bf tr}$ is the compensator field introduced there.

Eliminating the fields~$\Psi_{\bf div}$ and~$\Psi_{\Delta}$ using the second and fourth equations in~\eqn{pde},
we find that only the first and sixth of these equations are independent and give a description of all
massless, totally symmetric higher spins in terms of a pair of unconstrained fields~$(\Psi_{\bf grad},\Psi_{\bf tr})$
with a single unconstrained gauge parameter~$\xi$
\bea
{\bf G} \ \Psi_{\bf grad}&=&\frac 1 2 \ {\bf grad}^{3}\ \Psi_{\bf tr}\, ,\nn\\[2mm]
{\bf tr}^{2}\ \Psi_{\bf grad}&=&4\Big({\bf div}+\frac14{\bf grad \ tr}\Big)\ \Psi_{\bf tr}\, ,\nn\\[4mm]
\delta \Psi_{\bf grad}&=&{\bf grad}\ \xi \, ,\nn\\[3mm] \delta \Psi_{\bf tr}& =& {\bf tr}\ \xi\, .
\eea
Here the operator~${\bf G}$ is given by
\be
{\bf G}=\Delta-{\bf grad \ div} +\frac 12 \ {\bf grad}^{2 }\ {\bf tr}\, .
\ee
Although, for some contexts, a formulation of higher spin dynamics in terms of 
unconstrained fields can be useful, the above system has the disadvantage that it
has terms cubic in derivatives. This problem can be removed by using some
of the gauge freedom to set the field~$\Psi_{{\bf tr}}=0$.
This yields the standard description of massless, totally symmetric higher spins
in terms of a doubly trace-free symmetric tensor and a trace-free gauge parameter
\bea
{\bf G} \ \Psi_{\bf grad} = 0 = {\bf tr}^{2}\ \Psi_{\bf grad}\, ,\nn\\[2mm]
\delta \Psi_{\bf grad}={\bf grad}\ \xi \, ,\quad {\bf tr}\ \xi=0\, .
\eea
It is important to note that since we kept the rank~$s$, or in other words the
eigenvalue of the index operator~${\bf N}$, arbitrary these relations generate
the gauge invariant equations of motion for fields of any spin. We also remark that
the operators~$\{{\bf G},\ {\bf tr}^{2}\}$ themselves generate a first class algebra.

Finally, we can formulate this system in terms of a detour complex as follows.
The field equation~${\bf G}\ \Psi_{\bf grad}=0$ is equivalent to the equation
\be
{ \cal G}\  \Psi_{\bf grad}\equiv\Big(1-\frac14\  {\bf g\ tr}\Big)\ {\bf G}\  \Psi_{\bf grad}=0\, ,
\ee
where
\be
{\cal G} = \Delta-{\bf grad \ div} +\frac 12 \ \Big({\bf grad}^{2 }\ {\bf tr}+{\bf g} \ {\bf div}^{2}\Big)
-\frac14 \ {\bf g}\ \Big(2\Delta+{\bf grad\ div}\Big)\ {\bf tr}\,. \label{Einstein}
\ee
We will call this operator the higher spin Einstein operator, since if
$\Psi_{\grad} = h_{\mu\nu}dx^{\mu}dx^{\nu}$ 
it then produces the linearized Einstein tensor
\be
{\cal G} \Psi_{\bf grad} = (\Delta h_{\mu\nu}-2\nabla_{\mu}\nabla^{\rho}h_{\rho\nu}+\nabla_{\mu}\nabla_{\nu} h_{\rho}^{\rho}
+\eta_{\mu\nu}\nabla^{\rho}\nabla^{\sigma}h_{\rho\sigma}-\eta_{\mu\nu}\Delta h_{\rho}{}^{\rho} )dx^{\mu}dx^{\nu}\, .
\ee
The higher spin Einstein operator obeys identities
\bea
{\bf div}\ {\cal G} \ &=& 0\qquad {\rm mod}_{left} \;\;{\bf g}\, ,\label{bianchii}\\[1mm]
{\cal G}\ {\bf grad} &=& 0 \qquad {\rm mod}_{right} \;\; {\bf tr}\, ,\label{gaugei}
 \eea
where equality holds up to terms proportional to the operators~${\bf g}$ and~${\bf tr}$ acting from the left and right, respectively.
This allows us to form a detour complex
\be
\begin{array}{c}
0\longrightarrow 
\odot TM/_{\bf tr}
\stackrel{\bf grad} {\longrightarrow}
\odot TM/_{{\bf tr}^{2}}
\rightarrow\cdots
\qquad
\cdots\rightarrow 
\odot TM/_{{\bf tr}^{2}}
\stackrel{\bf div}{\longrightarrow} 
\odot TM/_{{\bf tr}^{2}}
\longrightarrow0
\\
\ \Big|\hspace{-.8mm}\raisebox{-2.5mm}{\underline{\quad\quad\quad\quad\ \ \ \ \ \raisebox{1mm}{${\cal G}$}\quad\quad\qquad\quad\; }} \hspace{-1.2mm}{\Big\uparrow}
\end{array}\label{DETOUR}
\ee
where~$\odot TM/\bullet$ denotes symmetric tensors modulo the relation~$\bullet$.
The operator~${\cal G}$ is formally self-adjoint so the equation of motion~${\cal G} \Psi_{\bf grad}=0$ comes from an action principle~$S=\frac12 \int (\Phi,{\cal G}\Phi)$ where the inner product~$(\cdot,\cdot)$ is the one inherited from the underlying quantum mechanical model. The relations~\eqn{bianchii}
and~\eqn{gaugei} express the Bianchi identity and gauge invariance of this field equation. 
Our next task is to generalize this construction to tensors of arbitrary symmetry types, and in particular
find compact expressions for the Einstein operators for these theories.

\subsection{Mixed  Tensors and $sp(2r)$ Quantum Mechanics}

Tensors transforming under arbitrary representations of~$gl(D)$ can be expressed
either in terms of:
\begin{enumerate}
\item
Tensors labeled by groups of antisymmetric indices~$\omega_{[\mu^{1}_{1}\ldots\mu^{1}_{k_{1}}]\cdots[\mu^{s}_{1}\ldots\mu^{s}_{k_{s}}]}$ ---``multi-forms''---
or schematically
\be
\Yvcentermath1
\yng(1,1,1,1)\; \otimes\; \yng(1,1,1)\; \otimes\cdots\otimes\; \yng(1,1,1,1,1)
\ee
where~$s$ labels the number of antisymmetric columns, while the~$k_{i}$ label the number of boxes in each column.
\item[{\it or}]
\item
Tensors labeled by groups of symmetric indices~$\varphi_{(\mu^{1}_{1}\ldots\mu^{1}_{s_{1}})\cdots(\mu^{r}_{1}\ldots\mu^{r}_{s_{r}})}$
 ---``multi-symmetric forms''--- or schematically
\be
\Yvcentermath1
\begin{array}{c}
\yng(5)\\\otimes\\ \yng(3)\\ \otimes\\\vdots\\\otimes\\ \yng(4)
\end{array}\label{symm}
\ee
where~$r$ labels the number of symmetric rows, while the~$s_{i}$ label the number of boxes in each row.
\end{enumerate}
In each case irreducible~$gl(D)$ representations are obtained by placing algebraic constraints on tensors of these types
akin to the Bianchi identity of the first kind obeyed by the Riemann tensor.
Supersymmetric quantum mechanical models whose Hilbert spaces are populated by the tensors described above were 
developed in~\cite{Hallowell:2007zb}. The~$R$ symmetry groups are~$O(2s)$ and~$Sp(2r)$ for the two respective cases. Models for tensors with both
symmetric and antisymmetric groups of indices also exist and have an~$osp(2s|2r)$~$R$-symmetry. The~${\cal N}=2$ and~$sp(2)$
(super)symmetric quantum mechanical models described above are the lowest lying examples of these.

Although, all the computations in this work, should in principle carry over to both~$osp(Q|2r)$ models for arbitrary
integers\footnote{Note that  when~$Q$ is odd, the model includes spinor fields~\cite{Hallowell:2007zb}.}~$(Q,r)$, here we concentrate on
the~$sp(2r)$ case. There are two reasons for this choice. Firstly, because the rows in~\eqn{symm} are symmetric, this allows us
to handle arbitrarily high spins without introducing arbitrarily quantum mechanical oscillator modes. In particular, if we take~$r\geq D$  
wavefunctions span tensors of arbitrary type. Secondly, since only bosonic 
oscillators are required, the BRST ghosts will all be fermionic leading to a BRST wavefunction with a finite expansion in ghost modes
and in turn a Lie algebra, rather than Lie superalgebra cohomology problem.

The quantum mechanical model whose wavefunctions describe  multi-symmetric forms derives from the simple action principle
\be
S=\int dt\left\{\frac12 \dot x^{\mu}\dot x_{\mu}+i z^{*}_{i\mu}\dot z^{i\mu}\right\}\, .\label{Ssp2k}
\ee
Here we have introduced~$2r$ oscillators ~$(z^{*}_{i\mu},z^{i\mu})$ with~$i=1\ldots r$. Their kinetic
term can be written in the manifestly~$sp(2r)$ symmetric way~$\frac i2 z_{\alpha\mu}\epsilon^{\alpha\beta}\dot z^{\mu}_{\beta}$
where~$z_{\alpha}=(z^*_{i},z^{i})$ and~$\epsilon^{\alpha\beta}$ is the antisymmetric invariant tensor of~$sp(2r)$.
Again, this theory can be coupled to curved backgrounds; we refer to~\cite{Hallowell:2007zb,Burkart:2008bq} for details.

Upon quantization the oscillators can be represented in terms of sets of commuting differentials~\cite{DuboisViolette:1999rd,Hallowell:2007zb}
\be
z_{i}^{*\mu}=d_{i}x^{\mu}\, ,\qquad z_{\mu}^{i}=\frac{\partial}{\partial(d_{i}x^{\mu})}\, .
\ee
The tensor depicted in~\eqn{symm} is then denoted
\be
\Phi=\varphi_{(\mu^{1}_{1}\ldots\mu^{1}_{s_{1}})\cdots(\mu^{k}_{1}\ldots\mu^{r}_{s_{r}})}d_{1}x^{\mu^{1}_{1}}\cdots d_{1}x^{\mu^{1}_{s_1}}
\; \cdots \;
d_{r}x^{\mu^{r}_{1}}\cdots d_{r}x^{\mu^{r}_{s_r}}\, .
\ee

The~$sp(2)$ generators~${\bf g}$,~${\bf N}$ and~${\bf tr}$ of the above Sections are promoted to~$r\times r$
matrices of operators:
\bea
{\bf g}_{ij}&=&d_{i}x^\mu g_{\mu\nu}d_{j}x^\nu\, ,\nn\\[4mm]
{\bf N}_{i}^{j}&=&dx_{i}^\mu \frac{\partial}{\partial (d_{j}x^{\mu})} ,\nn\\[2mm]
{\bf tr}^{ij}&=&\frac{\partial}{\partial (d_{i}x^{\mu})}g^{{\mu\nu}}\frac{\partial}{\partial (d_{j}x^{\nu})} \, .
\eea
The operators~$({\bf g}_{ij},{\bf N}_{i}^{j},{\bf tr}^{ij})$ generate~$sp(2r)$
\bea
[{\bf N}_{i}^{j},{\bf g}_{kl}]\, &=&\; \; \; 2\delta^{j}_{(k}{\bf g}_{l)i}\, ,\nn\\[2mm]
[{\bf tr}^{ij},{\bf g}_{kl}]\, &=&\;\, \ 4 \delta^{(i}_{(k}{\bf N}_{l)}^{j)}+2D\, \delta^{i}_{(k}\delta_{l)}^{j}\, ,\nn\\[2mm]
[{\bf N}_{i}^{j},{\bf tr}^{kl}] &=&-2\delta_{i}^{(k}{\bf tr}^{l)j}\, ,\nn\\[2mm]
[{\bf N}_{i}^{j},{\bf N}_{k}^{l}] &=& \; \; \; \ \delta_{k}^{j} {\bf N}_{i}^{l} - \delta_{i}^{l} {\bf N}_{j}^{k}\, .
\eea 
They correspond to ``$R$-symmetries'' of the model~\eqn{Ssp2k}. Geometrically, in terms of the picture~\eqn{symm},
they count the number of indices in a given row, move boxes from one row to
another, and add or remove pairs of boxes to or from (possibly distinct) rows using the metric tensor.

The differential operators~${\bf div}$ and~${\bf grad}$ are replaced by~$2r$ operators corresponding to the
divergence and gradient acting on each row in~\eqn{symm}. These form the fundamental representation of~$sp(2r)$.
\bea
[{\bf N}_{i}^{j},{\bf grad}_k]=\delta^{j}_{k}\ {\bf grad}_i\, ,\   && [{\bf N}_{i}^{j},{\bf div}^{k}]=-\delta_{i}^{k}\ {\bf div}^{j}\, ,\nn\\[2mm]
[{\bf tr}^{ij},{\bf grad}_k]=2\delta_{k}^{(i}\ {\bf div}^{j)}\, , && [{\bf g}_{ij},{\bf div}^{k}]=-2\delta_{(i}^{k}\ {\bf grad}_{j)}\, ,
\eea
and themselves obey the supersymmetry-like algebra
\bea
{}[\div^i,\grad_j]\ &=&\delta_{j}^{i}\ \Delta \, .
\eea
Before studying the Lie algebra cohomology of the above algebra,
and the accompanying spinning particle model, in the next Sections,
let us briefly discuss how irreducible tensor representations can be obtained
from the reducible ones depicted in~\eqn{symm}.

There are two pertinent notions of irreducibility for tensors. The first is with respect to~$gl(D)$
and is obtained by studying all possible permutation symmetries. This can be achieved using
the operators~$\N^{j}_{i}$ which move a box from row~$j$ to row~$i$ with a combinatorial
factor equaling the number of boxes in row~$j$, for example
\be
\N^{3}_{2}\left(
\begin{array}{c}
\yng(5)\\ \otimes\\[1mm] \yng(3)\\ \otimes \\[1mm] \yng(4)
\end{array}\right)
\; =\;
4\ \left(
\begin{array}{c}
\yng(5)\\ \otimes\\[1mm] \yng(4)\\ \otimes \\[1mm] \yng(3)
\end{array}\right)\, .
\ee
Irreducible~$gl(D)$ representations correspond to Young diagrams in which the number of 
boxes in each row decreases weakly (read from top to bottom). This amounts to tensors in the
kernel of all operators
\be
\{\N^{j>i}_{i}\}\, ,
\ee
which generate the nilradical subalgebra of~$gl(r)$.
 
For physical applications, often the stronger requirement of~$so(D)$ irreducibility is placed on tensors.
This amounts to additionally removing all traces and therefore tensors in the kernel of
\be
\{\N^{j>i}_{i},\ \tr^{ij}\}\, .\label{nil}
\ee
This set generates the nilradical of~$sp(2r)$. It plays an important {\it r\^ole} in the choice of first class 
algebra in the BRST construction of the next Section.

\subsection{Mixed Symmetry Einstein Operators}

\label{mixeinop}

The aim of the this Section is to derive the generalization of the Einstein operator~\eqn{Einstein}
to higher spins of arbitrary symmetry type. This result can actually be directly obtained by covariantizing~\eqn{Einstein}
with respect to\footnote{We thank Stanley Deser for this observation.}~$sp(2r)$  but here we outline its BRST derivation to connect with our path integral techniques. (For a review of existing available higher spin BRST techniques see~\cite{Fotopoulos:2008ka}, the unsymmetrized version of our Einstein operator was derived by BRST techniques recently in~\cite{Greg}. BRST quantization of the algebra~\eqn{exam2} below was also considered in~\cite{Tsu2}). 

To begin with we must choose the first class algebra generalizing~\eqn{that one}. The basic philosophy is that gauging~$\{\div^{i},\grad_{i}, \Delta\}$
yields the correct differential relations on propagating physical modes--this is perhaps seen most easily in the path integral approach described
in Section~\ref{Dofs}. We also expect algebraic relations that ensure the theory describes irreducible representations of the Lorentz group.
These will follow by also gauging the nilradical~$sp(2r)$ operators in~\eqn{nil}. Hence we study the Lie algebra cohomology of
\be
{\mathfrak g}=\Big\{\tr^{ij},\,  \div^{i},\ \N^{j>i}_{i},\, \grad_{i}, \, \Delta \Big\}\label{exam2}
\ee
acting on multi-symmetric forms. 

The next question we must address is what degree/ghost number cohomology corresponds to the underlying physical system.
This is resolved by recalling that the totally antisymmetric tensor theories possess gauge for gauge symmetries. For example, for 
a two-form~$\omega$ with gauge transformation~$\delta \omega = \bm d\alpha$, exact one-form gauge parameters~$\alpha = \bm d\beta$ do not 
act on~$\omega$. In other words, two-forms appear at degree two in de Rham cohomology. However, multi-symmetric forms with
$r$~rows as in~\eqn{symm} can also be expanded in multi-forms of degree~$r$ or less. This implies gauge for gauge symmetries
even for multi-symmetric forms and shows that we should study the degree~$r$ cohomology.

At degree~$r$ there are~$\binom{{\rm dim}{\mathfrak g}}{r} =
\binom{(r+1)^{2}}{r}$
closure relations on~$\binom{{\rm dim}{\mathfrak g}}{r-1}$  
fields (because here~${\rm dim}{\mathfrak g}=(r+1)^{2})$. 
These read
\be
{\mathfrak g}_{[A_{1}} \Psi_{A_{2}\ldots A_{r+1}]}- \frac r2 f^{B}_{[A_{1}A_{2}}\Psi_{|B|A_{3}\ldots A_{r+1}]}=0\, .\label{cl}
\ee
These fields enjoy~$\binom{{\rm dim \mathfrak g}}{r-1} =\binom{(r+1)^{2}}{r-1}$
gauge invariances following from exactness
\be
\delta \Psi_{A_{1}\ldots A_{{r}}}={\mathfrak g}_{[A_{1}} \xi_{A_{2}\ldots A_{r}]}- \frac{r-1}{2}f^{B}_{[A_{1}A_{2}}\xi_{|B|A_{3}\ldots A_{r}]}\, .\label{exactomundo}
\ee
Solving this system at arbitrary degree~$r$ may seem daunting, but is in fact not far more difficult than the~$r=1$,~$sp(2)$ computation
performed above. Let us sketch the main ideas and then give the result.

Firstly, we note that the field
\be
\Psi\equiv\Psi_{\grad_{1}\ldots \grad_{r}}=\frac 1 {r!} \epsilon^{i_{1}\ldots i_{r}}\Psi_{\grad_{i_{1}}\ldots\grad_{i_{r}}}
\ee 
gives the minimal covariant field content of our final detour complex. Here, and in what follows, we use a compact notation
where field labels~$\grad_{i}$ are soaked up with the~$sl(r)$ invariant, totally antisymmetric symbol
\be
\Psi_{\bullet}^{i_{1}\ldots i_{n}}\equiv \frac{1}{(r-n)!}\ \epsilon^{i_{1}\ldots i_{r}}\Psi_{\bullet \,  \grad_{i_{n+1}}\ldots \grad_{i_{r}}}\, .\label{epsilon}
\ee

Now we turn to the closure relations~\eqn{cl}.
When one of the adjoint indices is~$\Delta$ and all others are~$\grad_{i}$
we obtain a relation analogous to the last one in~\eqn{pde}
\be
\Delta\,  \Psi-\grad_{i}\ \Psi_{\Delta}^{\, i}=0\, .
\ee
The field~$\Psi_{\Delta}^{\, i}$ is not independent. It is eliminated by a pattern of closure relations analogous to the~$sp(2)$ ones in~\eqn{pde}
\bea
\div^{i}\,  \Psi - \grad_{j}\, \Psi^{\, j}_{\div^{i}}&=&\Psi_{\Delta}^{\, i}\, ,\nn\\[2mm]
\tr^{ij}\, \Psi - \grad_{r}\, \Psi^{\, r}_{\tr^{ij}}&=&\Psi^{j}_{\div^{i}}+\Psi^{i}_{\div^{j}}\, .\label{tortelloni}
\eea
These imply
\be
{\bf G} \, \Psi = \frac12 \, \grad_{i}\, \grad_{j}\, \grad_{k}\, \Psi^{k}_{\tr^{ij}}\, ,
\ee
where 
\be
{\bf G} = \Delta - \grad_{i}\div^{i}+\frac12\, \grad_{i}\grad_{j}\tr^{{ij}}\, .\label{quasieinstein}
\ee
(This operator was first derived by Labastida in~\cite{Labastida:1987kw}.)
As in the~$sp(2)$ case, the field~$\Psi^{k}_{\tr^{ij}}$ vanishes once we use the gauge freedom
implied by exactness. However, there are still double-trace relations satisfied by the physical field~$\Psi$. 
For these we consider the further closure relation
\be
\tr^{ij}\,\Psi_{\div^{k}}^{l}-\div^{k}\,\Psi_{\tr^{ij}}^{l}-\grad_{m}\,\Psi_{\div^{k}\tr^{ij}}^{ml}=
\Psi^{il}_{\div^{k}\div^{j}}+\Psi^{jl}_{\div^{k}\div^{i}}-\Psi_{\Delta\, \tr^{ij}}^{kl}\, .
\ee
This time the second and third terms on the left hand side can be gauged away as can the
antisymmetric part of~$\Psi^{i}_{\div^{j}}$ so, using~\eqn{tortelloni}, we must solve the equation
\be
\tr^{ij}\,\tr^{kl}\Psi=
\Psi^{il}_{\div^{k}\div^{j}}+\Psi^{jl}_{\div^{k}\div^{i}}-\Psi_{\Delta\, \tr^{ij}}^{kl}\, .
\ee
The antisymmetric part in~$k$ and~$l$ determines~$\Psi_{\Delta\, \tr^{ij}}^{kl}$ in terms of~$\Psi^{il}_{\div^{k}\div^{j}}$
which in turn depends on double traces of the physical field~$\Psi$. However, symmetrizing the above equation in
any three indices causes the right hand side to vanish identically. Therefore we learn the double trace relation
\be
\tr^{i(j}\tr^{{kl)}} \Psi = 0\, .
\ee
Finally we must remember the closure relations coming from gauging
the $R$-symmetries $\N^{j>i}_{i}$.  
In particular, using the trivial identity~$\delta^{{j>i}}_{i}=0$ we have
\be
\N^{j}_{i}\, \Psi - \grad_{k}\Psi^{k}_{\N^{j}_{i}}=0\, ,\qquad j>i\, .
\ee
Gauging away the second term leaves us with the algebraic constraint
\be
\N^{j>i}_{i}\, \Psi=0\, .
\ee
At this point there are no further constraints on the physical field~$\Psi$
and all remaining fields are either gauged away or algebraically dependent on~$\Psi$.
Let us gather together the equations of motion for~$\Psi$;
\be
{\bf G}\, \Psi = 0 = \tr^{i(j}\tr^{kl)}\, \Psi = \N^{j>i}_{i}\, \Psi\, ,\label{gather}
\ee
with~${\bf G}$ given in~\eqn{quasieinstein}.
As a consistency check, it is not difficult to verify that the operators
$\{{\bf G}$,~$\tr^{i(j}\tr^{kl)}$,~$ \N^{j>i}_{i}\}$ themselves form a first class algebra.

These equations of motion are gauge invariant under transformations
following from exactness~\eqn{exactomundo} 
\be
\delta \Psi = \frac 1{(r-1)!} \,  \epsilon^{i_{1}\ldots i_{r}}\grad_{i_{1}}\, \xi_{i_{2}\ldots i_{r}} \equiv \grad_{i} \xi^{i}\, .\label{gaugeo}
\ee 
In the last term we have employed the compact notation~\eqn{epsilon} also for the gauge parameters. The parameters
themselves are subject to constraints that can be deduced by carefully following which gauge freedoms were employed
to remove all independent fields save for the physical one~$\Psi$. These read
\be
\tr^{(ij}\, \xi^{k)}= \tr^{i(j}\tr^{kl)}\,\xi^m = 0 = \N^{j>i}_{i}\, \xi^{k}+\delta_i^k\,\xi^{j>i}\, .
\ee

The longhand notation in the second term of~\eqn{gaugeo} makes it clear that there are gauge for gauge symmetries
\be
\delta \xi_{\grad_{i_{1}}\ldots\grad_{i_{r-1}}}=\grad_{[i_{1}}\xi_{\grad_{i_{2}}\ldots\grad_{i_{r-2}]}}\, ,
\ee
coming from the Lie algebra cohomology at degree~$r-1$. Of course, there are further gauge for gauge symmetries
of the same form corresponding to the system being rank~$r$ when irreducible tensors are expanded in an  antisymmetric basis.

To end this Section, we compute the mixed symmetry analog of the
Einstein operator~\eqn{Einstein}. Firstly the field equation~${\bf
  G}\, \Psi=0$ is equivalent to
\be
{\cal G}\, \Psi \equiv \Big(1-\frac14\,  \g_{{ij}}\, \tr^{ij}+\frac1{48}\, \g_{ij}\g_{kl}\, \tr^{ij}\tr^{kl}\Big){\bf G}\Psi=0\, .
\ee 
(Here, and in the formula that follows we specialize to $r=2$ for simplicity). 
The mixed higher spin Einstein operator~${\cal G}$ then equals
\bea
{\cal G}&=&\Delta - \grad_{i}\, \div^{i}+\frac12\, \Big(\g_{ij}\,\div^{i}\div^{j}+\grad_{i}\grad_{j}\, \tr^{ij}\Big)\nn\\[3mm]
&-&\frac 14\, \g_{ij}\, \Big(2\Delta-\grad_{k}\, \div^{k}\Big)\, \tr^{ij}-\frac12\, \g_{ij}\grad_{k}\, \div^{i}\tr^{jk}\nn\\[3mm]
&+&\frac1{48}\, \g_{ij}\g_{kl}\Big(4\Delta+\grad_{m}\, \div^{m}\Big)\tr^{ij}\tr^{kl}\nn\\[3mm]
&-&\frac 1{6}\Big(\g_{i[j}\g_{k]l}\, \div^{i}\div^{j}\tr^{kl}+\g_{ij}\grad_{k}\grad_{l}\, \tr^{i[j}\tr^{k]l}\Big)\, .
\eea
It is self-adjoint and obeys Bianchi and gauge invariance identities
\bea
\div^{i}\, {\cal G}\, &=&0\qquad \mbox{mod}_{left} \; \g_{ij}\, ,\\
{\cal G}\,\grad_{i} &=&0\qquad \mbox{mod}_{right} \; \tr^{ij}\, ,
\eea
ensuring the existence of a detour complex analogous to~\eqn{DETOUR}. Once again, an action principle~$S=\frac12 \int (\Phi,{\cal G}\Phi)$ with inner product~$(\cdot,\cdot)$ inherited from the underlying quantum mechanics also follows immediately.
Our next task is to quantize this system. We adopt a first quantized
approach which we now explain.

\subsection{The $sp(2r)$ Spinning Particle}

\label{sp2rspin}

The particle action introduced in the previous Section in
equation~(\ref{Ssp2k}) is invariant under global extended supersymmetry, with
$Sp(2r)$ R-symmetry group. This action can be used 
to construct locally supersymmetric particle actions that give a
path integral implementation of the quantum algebras studied
previously. 

Let us start from the phase space symplectic integral
\bea
S= \int dt \left\{p_\mu \dot x^\mu +iz_{i\mu}^* \dot z^{i\mu}\right\}\, ,
\eea
which is invariant under the action of the global transformations with symmetry generator
\bea
G = \xi H + \bar \sigma_i S^i + \sigma^i \bar S_i +\frac12\beta^{ij}
\bar K_{ij} +\frac12\bar \beta_{ij} K^{ij}+\alpha^i_j J_i^j\, ,
\eea 
where the classical susy generators and classical~$sp(2r)$ generators are,
respectively, given by
\bea
&& \bar S_i = p \cdot z_i^*\,,\quad S^i = p \cdot
z^i\,,\quad H = \frac12\,  p^2\, ,\\[2mm]
&& \bar K_{ij} = z_i^* \cdot z_j^*\, ,\quad 
J_{i}^{j} =z_i^*\cdot z^j\,, \quad K^{ij} =z^i\cdot z^j\, .
\eea
Here the~$u(r)$ subalgebra generated by~$J_{i}^{j}$ is made manifest. Note that the canonical
quantization discussed earlier simply amounts to the replacement
\bea
&&i \bar S_i \to {\bf grad}_i\,,\quad i S^i \to {\bf div}^i\,,\quad -2H
\to \Delta \nn\\[2mm]
&& \bar K_{ij} \to {\bf g}_{ij}\, ,\quad 
J_{i}^{j} \to {\bf N}_i^j\,, \quad K^{ij}\to {\bf tr}^{ij}~.
\eea
The transformation rules for the dynamical fields can be read off from
$\delta q = \left\{q,G\right\}$
\bea
\delta x^\mu &=& \xi\, p^\mu +\bar\sigma_i\, z^{i \mu} +\sigma^i\, z_i^{* \mu}\, ,\nn\\[1mm]
\delta p_\mu &=& 0\, ,\nn\\[1mm]
\delta z^{i \mu} &=& -i\sigma^i\, p^\mu-i\alpha^i_j\, z^{j\mu}
-i\beta^{ij}\, z_j^{*\mu}\, ,\nn\\[1mm]
\delta z^*_{i \mu} &=&  i\bar \sigma_i\, p^\mu+i\alpha^j_i\, z^*_{j\mu}
+i\bar \beta_{ij}\, z^{j}_\mu~.
\eea 
Gauged actions are
thus obtained by adding gauge fields coupled to the above conserved
charges. In particular, gauging all the global symmetries yields the
action  
\be
S= \int dt \left\{p_\mu \dot x^\mu +iz_{i\mu}^* \dot
z^{i\mu}-eH - \bar s_i S^i-s^i \bar S_i-\frac12 \bar b_{ij} K^{ij}
-\frac12 b^{ij} \bar K_{ij} - a^i_j J_i^j\right\}\, ,
 \ee 
and the transformations for the gauge fields are obtained by
requiring~$S$ to be invariant under local transformations~$\Xi(t)=(\xi(t),
\bar \sigma_i(t),\sigma^i(t), \bar \beta_{ij}(t),\beta^{ij}(t),\alpha^i_j(t))$, namely
\bea
\delta e \;\  &=& \dot\xi+i2\sigma^i \bar s_i-i2 \bar\sigma_i s^i\, ,\nn\\[1mm]
\delta s^i \ &=& \dot \sigma^i +i a^i_j \sigma^j-i\alpha^i_j
s^j+i\beta^{ij}\bar s_j -ib^{ij}\bar \sigma_j\, ,\nn\\[1mm]
\delta \bar s_i\  &=& \dot {\bar {\sigma_i}}-i a^j_i \bar \sigma_j+i\alpha^j_i
\bar s_j-i\bar \beta_{ij} s^j +i\bar b_{ij}\sigma^j\, , \nn\\[2mm]
\delta b^{ij} &=& \dot \beta^{ij}+i\left(a^i_k\,
\beta^{jk}+a^j_k\, \beta^{ik}\right)
-i\left(\alpha^i_k\, b^{jk}+\alpha^j_k\, b^{ik}\right)\, ,\nn\\[1mm]
\delta \bar b_{ij} &=&\!\!  \dot {\bar{\,\,\beta_{ij}}}-i\left(a^k_i\,
\bar\beta_{jk}+a^k_j\, \bar\beta_{ik}\right)
+i\left(\alpha^k_i\, \bar b_{jk}+\alpha^k_j\, \bar b_{ik}\right)\, ,\nn\\[1mm]
\delta a^i_j &=& \dot \alpha^i_j-i\left(\alpha^i_k\,
a^k_j-a^i_k\,\alpha^k_j\right) +i\left(\beta^{ik}\,\bar b_{jk}
-b^{ik}\bar\beta_{jk}\right)~. 
\eea
However, it will be most interesting to consider partial gaugings that
only involve subalgebras of the above algebras. In fact, in many of these cases
it is possible to leave (part of) the abelian subgroup~$U(1)^r \subset U(r)$
invariant, which would allow a gauge-invariant Chern-Simons action
\bea
S_{{\rm CS}} = \int dt \sum_i q_i\ a^i_i~.
\eea  
This can be used to fix the number of indices in a particular row of a Young diagram.

Let us single out a few interesting cases   
\begin{enumerate}
\item \underline{\it Gauge~$H,\ \bar S_i,\ S^i,\ K^{ij},\ J^i_j$} (or $\Delta$, ${\bf grad}$, ${\bf div}$, ${\bf tr}$ and {\it all} ${\bf N}$'s). 
This amounts to setting~$b^{ij} = \beta^{ij}=0$ in the
previous transformations rules and leaves in particular
\bea
\delta a^i_j &=& \dot \alpha^i_j-i\left(\alpha^i_k\,
a^k_j-a^i_k\,\alpha^k_j\right)
\eea
from which it is obvious that~$a =\sum_i^r a^i_i$ is invariant and the unique Chern--Simons term is
$S_{{\rm CS}} =\int dt\ q\, a$.
\item \underline{\it Gauge~$H,\ \bar S_i,\ S^i,\ K^{ij},\ J^{j\geq i}_i$} (or $\Delta$, ${\bf grad}$, ${\bf div}$, ${\bf tr}$ and nilradical ${\bf N}$'s). This
  amounts to setting~$b^{ij} =\beta^{ij}=0$ and~$a^i_j
  =\alpha^i_j=0$ when~$i >j$, from which
\bea
\delta a^i_i &=& 0\,,\quad {\rm (no\ sum\ implied)}
\eea
and allowed Chern--Simons are given by $S_{{\rm CS}} =\int dt\ \sum_i q_i\, a^i_i$.

\end{enumerate}
We are now ready to quantize these gauged models.

\subsection{Counting Degrees of Freedom}\label{Dofs}

In the present Section we use the particle actions described above to
compute the number of degrees of freedom for mixed higher spin tensor
multiplets. We study the partition function 
\bea
Z\sim \int_{S^1} \frac{DX DE}{\rm Vol (Gauge)}\, e^{iS[X,E]+iS_{{\rm CS}}[E]} \, ,
\eea 
where~$X$ collectively denotes all dynamical fields, whereas~$E$
denotes gauge fields. In the latter we need to carefully gauge fix all the
gauge symmetries present in the spinning particle action. We now Wick rotate
to Euclidean time (periodic boundary conditions) and 
use the
Faddeev-Popov trick to extract the volume of the gauge group to set a
gauge choice that completely fixes all the supergravity fields up to
some constant moduli fields
\be E=(e,\bar s_i, s^i, \bar
b_{ij},b^{ij},a^i{}_j)=(\beta,0,0,0,0,\theta_i\delta^i_j)\, ,\ee
where~$\hat a^i{}_j =\theta_i\delta^i_j$ is the most generic constant element
of the Cartan subalgebra of~$sp(2r)$, with~$\theta_i$ being angles
taking values in a fundamental domain. We thus have
\bea
Z =-\frac12 \int_0^\infty \frac{d\beta}{\beta}\int
\frac{d^Dx}{(2\pi\beta)^{D/2}}\ {\rm DoF}(q,r)
\eea
with
\bea  
{\rm DoF}(q,r) &\equiv&  
K_r\hskip-.1cm \underbrace{ 
\prod_{i=1}^r \bigg [ 
\int_0^{2\pi} \frac{d\theta_i}{2\pi }\, e^{iq_i \theta_i} \bigg]}_{ 
{\rm Cartan \, moduli\, for\, }\, Sp(2r)\, +\, {\rm CS}} 
\underbrace{ 
\prod_{i=1}^r \bigg [  
{\rm Det}(\partial_\tau -i\theta_i) 
\bigg ] 
}_{\bar S_i} \ 
\underbrace{ 
\prod_{i=1}^r \bigg [  
{\rm Det}(\partial_\tau +i\theta_i) 
\bigg ] 
}_{S^i}  \nn\\ 
&\times& \underbrace{ 
\prod_{i=1}^r \bigg [  
{\rm Det}(\partial_\tau +i\theta_i)]^{-D} \bigg ]}_{z,z^*} \  
\underbrace{\prod_{i=1}^r \bigg [  
{\rm Det}(\partial_\tau -2i\theta_i)\bigg ] 
}_{\bar K_{ii}\ {\rm no\ sum}}  \ 
\underbrace{\prod_{i=1}^r \bigg [  
{\rm Det}(\partial_\tau +2i\theta_i)\bigg ] 
}_{K^{ii}\ {\rm no\ sum}} 
\nn\\ 
&\times& 
\underbrace{ 
\prod_{i\neq j} \bigg [  
{\rm Det} \Big(\partial_\tau -i(\theta_i-\theta_j)\Big) 
\bigg ]\ 
}_{U(r) \, {\rm step\ operator}}
\underbrace{\prod_{i<j} \bigg [  
{\rm Det} \Big (\partial_\tau -i(\theta_i+\theta_j)\Big) 
\bigg ]}_{\bar K_{ij},\ i\neq j}  \nn\\
&\times&   
\underbrace{\prod_{i<j} \bigg [  
{\rm Det} \Big (\partial_\tau +i(\theta_i+\theta_j)\Big) 
\bigg ]}_{K^{ij},\ i\neq j}   
\eea  
being the number of degrees of freedom, and~$K^{-1}_r$ the
number of fundamental domains included in the integration domain. All
the determinants are evaluated with periodic boundary conditions because the
fields traced over are bosonic. 

The latter expression should really 
be understood as a generating function, including all ingredients for all
possible~$sp(2r)$ particle actions. Clearly, for each specific gauged action, one
has to pick out only those determinants that are involved in its gauge
fixing. Let us consider the example~2 of Section~\ref{sp2rspin}, which we claim corresponds to the 
BRST cohomology of the algebra~\eqn{exam2} computed in Section~\ref{mixeinop}.
For
that  theory we clearly have
\bea  
{\rm DoF}(q,r) &=&  
K_r\ \prod_{i=1}^r \bigg [ 
\int_0^{2\pi} \frac{d\theta_i}{2\pi }\, e^{iq_i \theta_i} \bigg]
\prod_{i=1}^r \bigg [  
{\rm Det}(\partial_\tau +i\theta_i)]^{2-D} \bigg ]  
\prod_{i=1}^r \bigg [  
{\rm Det}(\partial_\tau -2i\theta_i)\bigg ]   
\nn\\ 
&\times& 
\prod_{i< j} \bigg [  
{\rm Det} \Big(\partial_\tau -i(\theta_j-\theta_i)\Big) 
\bigg ]\ 
\prod_{i<j} \bigg [  
{\rm Det} \Big (\partial_\tau +i(\theta_i+\theta_j)\Big) 
\bigg ]~.\eea
Using that
\bea
{\rm Det}(\partial_\tau +i\theta) =2i \sin (\theta/2)
\eea
and making the change of variables~$w_i = e^{-i\theta_i}$ we obtain
\be
{\rm DoF}(q,r) =
K_r\, \prod_{i=1}^r \bigg [- 
\oint \frac{dw_i}{2\pi i w_i }\,
\frac{(1-w_i)^{2-D}(w_i^2-1)}{w_i^{q_i+r+1-D/2}}\bigg]
\prod_{i< j}(w_j-w_i)(1-w_jw_i)
\ee  
that gives
\bea
{\rm DoF}(q,r)
&=&  
K_r\, \prod_{i=1}^r\bigg[- \frac1{n_i!}\frac{d^{n_i}}{dw_i^{n_i}}\bigg]\nn\\
&\times&\prod_{i=1}^r (1-w_i)^{2-D}(w_i^2-1)
\prod_{i< j}(w_j-w_i)(1-w_jw_i)\Big|_{w_i=0}
\label{dofs}
\eea   
with~$n_i = q_i+r+1-D/2$~. 

Let us consider a few specific examples. The
simplest one is clearly~$r=1$ for which we set~$s\equiv q+2-D/2=n$
and, using that~$K_1=1$, we obtain
\bea
{\rm DoF}(s,1) = \frac{D+2s-4}{s}\left(
\begin{array}{c}
D+s-5\\
s-1
\end{array}
\right)
\eea
which  is precisely the dimension of a Young tableau
of~$so(D-2)$ with~$1$ row and~$s$ columns.  

For~$r=2$, we identify~$n_2\equiv s_2 +1$ and~$n_1\equiv s_1$, then $K_2=1$ and
using~(\ref{dofs}) we obtain
\bea
{\rm DoF}(s_2,s_1,2) &=& \frac{(D+s_1-7)!
  (D+s_2-6)!(s_2-s_1+1)!}
{(D-6)!(D-4)!(s_2+1)!\, s_1!\, (s_2-s_1)!}\nn\\[1mm]
&\times& (D+s_2+s_1-5)(D+2s_2-4)(D+2s_1-6)\, .
\eea 
This is the dimension of a Young tableau with~$s_2$ boxes in the
first row and~$s_1\leq s_2$ boxes in the second row. 

For arbitrary~$r$ we  identify~$n_k=s_k+k-1$ so 
that~(\ref{dofs}) should yield the dimension of a generic Young tableau with   
$s_k$ boxes in the~$k-$th row and~$s_1\leq s_2\leq \cdots \leq
s_r$. In all these cases, the physical degrees of freedom correspond then to
irreducible~$so(D-2)$ representations. 
In fact, for arbitrary $r$, we expect that equation~\eqn{dofs} is the generating function for the dimensions
of irreducible ~$so(D-2)$ 
representations. 
In the next Section we obtain the same result from the second quantized equations
of motion that follow from the BRST cohomology computation of Section~\ref{mixeinop}.

\subsection{Lightcone Degrees of Freedom}

Our final computation is to verify that the path integral degree of freedom counts 
match those obtained by a direct analysis of the second quantized
field equations \eqn{gather} which we reproduce here for convenience
\bea
 \Big(\Delta - \grad_{i}\div^{i}+\frac12\, \grad_{i}\grad_{j}\tr^{{ij}}\Big) \Psi &=& 0\, ,\label{feq}\\[1mm]
 \tr^{i(j}\tr^{kl)}\, \Psi & =&0\, ,\\[2mm]
  \N^{j>i}_{i}\, \Psi&=&0\, .\label{symmetry}
\eea
These equations enjoy the gauge invariances
\be
\delta\Psi=\grad_{i}\xi^{i}\, ,\quad\mbox{where }
\tr^{(ij} \, \xi^{k)}=\tr^{i(j} \tr^{kl)}\, \xi^{m}= 0\, .\label{gau}
\ee

A detailed proof of these dimension counts, which are essentially just a higher dimensional analog of Wigner's original computation of unitary representations of the Poincar\'e group~\cite{Wigner}, were only given rather recently in~\cite{Bekaert:2006ix}. A BRST lightcone version of this computation
was given in~\cite{Greg}.
By far the the speediest method to perform this computation, however, is to employ lightcone gauge directly to the second quantized
Lagrangian. For completeness we present that result here.  Expressing the metric as
\be
ds^{2}=2dx^{+} dx^{-} + d\vec x{\, }^{2}\, ,
\ee
we assume that~$\partial\slash \partial x^{-}$ is invertible and set it equal to unity in what follows.

Our main philosophy is to expand fields and field equations in powers of differentials~$d_{i}x^{-}$ in the~$x^{-}$ direction.
All the operators~$(\Delta,\grad_{i},\div^{i},\g_{ij},\N^{j}_{i},\tr^{ij})$ then have~$(D-2)$-dimensional analogs operating in the
$\vec x$ directions. We denote these by hats so that
\bea
\Delta\ \ &=&2\, \frac\partial{\partial x^{+}}\ +\ \widehat \Delta\, ,\nn\\[2mm]
\grad_{i} &=& d_{i }x^{-}+d_{i}x^{+}\frac\partial{\partial x^{+}}+\widehat\grad_{i}\, ,\nn\\[2mm]
\div^{i}\  &=& \frac\partial{\partial(d_{i }x^{-})}\, \frac\partial{\partial x^{+}}+\frac\partial{\partial(d_{i }x^{+})}+\widehat\div{}^{i}\, ,\nn\\[2mm]
\g_{ij}\ &=& 2\, d_{(i}x^{-}d_{j)}x^{+}+\widehat\g_{{ij}}\, ,\nn\\[2mm]
\N_{i}^{j}\ &=& d_{i}x^{-} \frac\partial{\partial(d_{j }x^{-})}+d_{i}x^{+} \frac\partial{\partial(d_{j }x^{+})}+\widehat\N_{i}^{j}\, ,\nn\\[2mm]
\tr^{{ij}}&=&2\frac\partial{\partial(d_{(i }x^{-})}\frac\partial{\partial(d_{j) }x^{+})} +\widehat\tr{}^{ij}\, .\label{deco}
\eea
Now we decompose the field~$\Psi$ as
\be
\Psi(d_{i}x^{-})=\psi + d_{i}x^{-}\chi^{i}\, ,
\ee
where~$\psi$ is independent of~$d_{i}x^{-}$.
On the other hand the fields~$\chi^{i}$ are~$d_{i}x^{-}$ dependent, and we focus on the term with the highest power of~$d_{i}x^{-}$.
Examining the terms of highest order in the field equation~\eqn{feq} coming from the~$\grad^{2}\, \tr$-term, we see that the highest order term
in~$\chi^{i}$ is lightcone symmetric-trace-free ({\it i.e.}, annihilated by~$\widehat\tr{}^{(ij}$).
However, from the lightcone decomposition of~$\grad_{i}$ in~\eqn{deco}, the gauge invariance~\eqn{gau} becomes
\be
\delta \Psi = \Big(d_{i}x^{-}+\cdots \Big) \xi^{i}\, .
\ee
The trace condition on the gauge parameter~$\xi^{i}$ exactly ensures that its highest~$d_{i}x^{-}$ term is also lightcone symmetric-trace-free.
Hence we may algebraically gauge away the highest order term in~$\chi^{i}$. Iterating the above argument allows us to gauge away all of 
$\chi^{i}$ so that
\be
\Psi=\psi(d_{i}\vec x,d_{i}x^{+})\, .
\ee
Our computation is completed by solving~\eqn{feq} for fields of the above form. Again we work order by order in~$d_{i}x^{-}$.
At highest order we learn
\be
\widehat \tr{}^{ij}\, \psi=0\, .
\ee
To study lower order terms we split
\be
\psi(d_{i}\vec x,d_{i} x^{+})=\widehat \psi(d_{i}\vec x) + d_i x^{+} \psi^{i}\, ,\label{Xpand}
\ee
where the field~$\widehat \psi$ only has~$(D-2)$-dimensional indices.
The fields~$\psi^{i}$ are all dependent because the next to leading order terms in~\eqn{feq} imply
\be
\frac{\partial\psi}{\partial (d_{i}x^{+})} + \widehat\div{}_{i}\psi=0\, .
\ee
This condition can always be solved in terms of the~$\psi_{i}$ in~\eqn{Xpand} so it remains to 
gather the remaining lowest order terms in~\eqn{feq} which read
\be
\Big\{ 
2\frac{\partial}{\partial x^{+}}+\widehat \Delta\Big\}\, 
\widehat \psi = 0\, .
\ee
This is simply the~$D$-dimensional Klein--Gordon equation. Finally we still need to impose the symmetry condition.
It is not hard to see that it implies
\be
\widehat \N_{i}^{j>i}\, \widehat \psi = 0\, .
\ee
Hence the independent light cone degree of freedom are described by a totally symmetric
$(D-2)$-dimensional tensors, which solve the~$D$-dimensional wave equation and are both~$(D-2)$-dimensional trace-free and irreducible.
Or in other words, the degree of freedom count is given by dimensions of~$so(D-2)$ irreducible representations.
This shows the claimed equivalence between BRST and path integral quantizations.

\section{Conclusions}\label{conclusions}

In this Article we have tackled the problem of constructing and quantizing quantum field theories for tensor fields 
with general symmetry types using a worldline approach. As depicted below, our starting point was a quantum mechanical 
(super)symmetric model whose wave functions are the type of tensor fields appearing in the desired second quantized model.
\begin{center}
\begin{tabular}{ccccc}
\\
&&(Super)symmetric&&\\[1mm] &&Quantum Mechanics&& \\[2mm]
&\scalebox{2}{$\swarrow$}&&\scalebox{2}{$\searrow$}\\[2mm]
Spinning Particle &&
&&
BRST Detour\\[1mm] Path Integral&&\scalebox{2}{$\longleftarrow\!\!\longrightarrow$}&&Quantization
\end{tabular}
\end{center}
Thereafter we identified first class constraint operator algebras acting on the quantum mechanical Hilbert space.
From these algebras one can build a first quantized gauge theory in two ways, either path integral methods, or
BRST quantization. The former led to a path integral representation of a spinning particle model
while the latter, using the detour complex idea, yielded the classical equations of motion of a second quantized
gauge field theory. The path integral approach gave a worldline method for computing quantum quantities
in the second quantized field theory, the simplest of which was a count of the physical degrees of freedom.
These can of course also be computed by studying the dimensionality of the Cauchy data of the classical field equations.
Indeed we found that these two methods gave identical answers for a large class of higher spin theories.

The quantum mechanical models we used as a starting point fall into a very broad class of models labeled by their
$R$-symmetry groups which are given by general orthosymplectic supergroups. In that language our Article focused
on the~$osp(2|0)$ and~$osp(0|2r)$ models. However, it is clear that our methods can be generalized to any of the~$osp(Q|2r)$ models.
When~$Q$ is odd, these models describe spinor-tensor fields in second quantization. The case~$osp(2|1)$ has been studied in~\cite{Hallowell:2005np}
to describe spinor valued totally symmetric tensor theories, but clearly a complete description of fermionic second quantized models
would be desirable.

There are many other directions our results lead to. The most interesting of course, would be to shed light on self-interactions
of higher spin fields. By now a large literature exists on this subject, a consistent theme being that interactions
for higher spins requires towers of infinitely many second quantized fields (see~\cite{Bekaert:2005vh} for an extensive review of these
developments). In simplest terms this points at a difficulty gauging the number operator(s)~${\bf N}$ of our quantum mechanical 
models. From the path integral viewpoint, this difficulty can be seen through the sparsity of consistent world-line
Chern--Simons terms that can be added to the worldline action.

There are two other most interesting, and in fact related, applications of our results. These are computations of higher
(second quantized) quantum amplitudes and interactions with backgrounds. Higher amplitudes are encoded,
for example, by studying the dependence of the worldline effective action on {\it arbitrary} background fields. It is not difficult to couple our
underlying quantum mechanical models to either background Yang--Mills or gravitational fields by twisting the connection
appearing in the covariant canonical momentum operator. However in general this can produce obstructions to our constraint
algebras being first class. These obstructions have been studied and explicated in~\cite{Burkart:2008bq}. The phenomenon of higher spin fields suffering inconsistencies in backgrounds is one that has been known for a long time
(dating back to work on coupling massive spin 3/2 fields, see for a thorough account~\cite{Deser:2001dt}).

It is possible to view these obstructions to first class algebras in general spaces in a more positive light. Namely,
these algebras can be used to develop powerful invariants for determining the underlying geometry of the 
background manifold. In turn, when the background manifold belongs to a special class of geometries, 
consistency and even enhanced symmetries and constraint algebras can result. The special {\it r\^ole} played
by certain geometries in string theory is an example of this phenomenon. Another example are the K\"ahler
higher spin models constructed in~\cite{Marcus:1994em,Marcus:1994mm,Bastianelli:2009vj} and~$(p,q)$-form K\"ahler electromagnetism~\cite{Cherney:2009vg}. The latter of these
theories follows from the detour construction~\cite{CLW}. It would be most interesting to compute its path integral quantization. 

\section*{Acknowledgments}

A.W. thanks INFN and the University of Bologna for the warmest hospitality during a visit in which 
the bulk of this work was completed. We thank Stanley Deser, Emanuele Latini, Andy Neitzke and Boris
Pioline for collaborations during early stages of this work. The work
of F.B. and O.C. was partly supported by the Italian MIUR-PRIN
contract 20075ATT78.

\end{document}